\documentclass{aastex61}

\shorttitle{Optical Monitoring S5 0716+714}
\shortauthors{Xiong et al.}

\begin{document}
\title{Multicolor Optical Monitoring of the Blazar S5 0716+714 from 2017 to 2019}

\author{Dingrong Xiong}
\author{Jinming Bai}
\affil{Yunnan Observatories,
Chinese Academy of Sciences, 396 Yangfangwang, Guandu District, Kunming, 650216, P. R. China; xiongdingrong@ynao.ac.cn; baijinming@ynao.ac.cn}
\affil{Center for Astronomical Mega-Science, Chinese Academy of Sciences, 20A Datun Road, Chaoyang District, Beijing, 100012, P. R. China}
\affil{Key Laboratory for the Structure and Evolution of Celestial
Objects, Chinese Academy of Sciences, 396 Yangfangwang, Guandu District, Kunming, 650216, P. R. China}
\author{Junhui Fan}
\affil{Center for Astrophysics, Guangzhou University, Guangzhou 510006, P. R. China}
\affil{Astronomy Science and Technology Research Laboratory of Department of Education of Guangdong Province, Guangzhou 510006, P. R. China}
\author{Dahai Yan}
\affil{Yunnan Observatories,
Chinese Academy of Sciences, 396 Yangfangwang, Guandu District, Kunming, 650216, P. R. China; xiongdingrong@ynao.ac.cn; baijinming@ynao.ac.cn}
\affil{Center for Astronomical Mega-Science, Chinese Academy of Sciences, 20A Datun Road, Chaoyang District, Beijing, 100012, P. R. China}
\affil{Key Laboratory for the Structure and Evolution of Celestial
Objects, Chinese Academy of Sciences, 396 Yangfangwang, Guandu District, Kunming, 650216, P. R. China}
\author{Minfeng Gu}
\affil{Key Laboratory for Research in Galaxies and Cosmology, Shanghai Astronomical Observatory, Chinese Academy of Sciences, 80 Nandan Road, Shanghai 200030, P. R. China}
\author{Xuliang Fan}
\affil{School of Mathematics, Physics and Statistics, Shanghai University of Engineering Science, Shanghai 201620, P. R. China}
\author{Jirong Mao}
\affil{Yunnan Observatories,
Chinese Academy of Sciences, 396 Yangfangwang, Guandu District, Kunming, 650216, P. R. China; xiongdingrong@ynao.ac.cn; baijinming@ynao.ac.cn}
\affil{Center for Astronomical Mega-Science, Chinese Academy of Sciences, 20A Datun Road, Chaoyang District, Beijing, 100012, P. R. China}
\affil{Key Laboratory for the Structure and Evolution of Celestial
Objects, Chinese Academy of Sciences, 396 Yangfangwang, Guandu District, Kunming, 650216, P. R. China}
\author{Nan Ding}
\affil{School of Astronomy and Space Science, Nanjing University, Nanjing, Jiangsu 210093, P. R. China}
\affil{Key Laboratory of Modern Astronomy and Astrophysics (Nanjing University), Ministry of Education, Nanjing 210093, P. R. China}
\author{Rui Xue}
\affil{School of Astronomy and Space Science, Nanjing University, Nanjing, Jiangsu 210093, P. R. China}
\affil{Key Laboratory of Modern Astronomy and Astrophysics (Nanjing University), Ministry of Education, Nanjing 210093, P. R. China}
\author{Weimin Yi}
\affil{Yunnan Observatories,
Chinese Academy of Sciences, 396 Yangfangwang, Guandu District, Kunming, 650216, P. R. China; xiongdingrong@ynao.ac.cn; baijinming@ynao.ac.cn}
\affil{Center for Astronomical Mega-Science, Chinese Academy of Sciences, 20A Datun Road, Chaoyang District, Beijing, 100012, P. R. China}
\affil{Key Laboratory for the Structure and Evolution of Celestial
Objects, Chinese Academy of Sciences, 396 Yangfangwang, Guandu District, Kunming, 650216, P. R. China}
\affil{Department of Astronomy and Astrophysics, The Pennsylvania State University, 525 Davey Lab, University Park, PA 16802, USA}

\begin{abstract}
We continuously monitored the blazar S5 0716+714 in the optical $g$, $r$ and $i$ bands from Nov. 10, 2017 to Jun. 06, 2019. The total number of observations is 201 nights including 26973 data points. This is a very large quasi-simultaneous multicolor sample for the blazar. The average time spans and time resolutions are 3.4 hours and 2.9 minutes per night, respectively. During the period of observations, the target source in the $r$ band brightens from $14^{\rm m}.16$ to $12^{\rm m}.29$ together with five prominent sub-flares, and then first becomes fainter to $14^{\rm m}.76$ and again brightens to $12^{\rm m}.94$ with seven prominent sub-flares. For the long-term variations, we find a strong flatter when brighter (FWB) trend at a low flux state and then a weak FWB trend at a higher flux state. A weak FWB trend at a low flux state and then a strong FWB trend at a higher flux state are also reported. Most of sub-flares show the strong FWB trends, except for two flares with a weak FWB trend. The particle acceleration and cooling mechanisms together with the superposition of different FWB-slopes from sub-flares are likely to explain the optical color behaviours. A scenario of bent jet is discussed.
\end{abstract}


\section{INTRODUCTION}

Blazars, the most energetic sustained sources in the nature, are a special kind of radio-loud active galactic nuclei (AGN), including one of relativistic jets oriented at a small angle with respect to the line of sight (Urry \& Padovani 1995; Padovani et al. 2016). Many physical properties from blazars can be explained by the small viewing angle between jet direction and the line of sight (Blandford \& Rees 1978). The small viewing angle produces a Doppler beaming effect that foreshortens the observed time-scales, blue-shifts the observed spectrum and magnifies the luminosities at all wavelengths (Falomo et al. 2014). The non-thermal emission from blazars has been detected in very wide wavebands from the radio band to the TeV energies. They are the most dominant high/very high-energy gamma-ray sources (Fermi-LAT Collaboration et al. 2019; TeVCat\footnote{http://tevcat.uchicago.edu/}). The IceCube Collaboration and multi-messenger observations had detected a very high energy neutrino from the direction of the BL Lac object TXS 0506+056 at a 3-3.5$\sigma$ level (IceCube Collaboration et al. 2018a; also see IceCube Collaboration et al. 2018b for the 2014-2015 neutrino flare). So blazars have offered a unique opportunity to explore the jet properties, high/very high-energy gamma-rays, neutrinos and cosmic rays (e.g., Murase et al. 2014, 2018).

Traditionally, blazars are divided into two categories, namely BL Lacertae (BL Lac) objects and flat spectrum radio quasars (FSRQs), according to the differences of equivalent width (EW) of the emission lines (Urry \& Padovani 1995). This classification criterion may lead to misclassification of blazars because the EW is related with optical continuum (Ghisellini et al. 2011). The optical continuum is related with the flux state. A small EW may be caused by the enhancement of optical continuum rather than low luminosity of emission lines. In a very low flux state, a high EW may be due to low luminosity of optical continuum for a source of intrinsically weak lines. A more physical criterion based on the $L_{\rm BLR}/L_{\rm Edd}$ is proposed (Ghisellini et al. 2011; Xiong \& Zhang 2014). BL Lac objects have very weak or absent emission lines while FSRQs have strong emission lines. BL Lac objects offer a more direct view into the emission from relativistic jet due to the lack of a substantial thermal component. Two prominent peaks appear in the broadband spectral energy distributions (SEDs) in which the emission from the low energy peak (IR-X-ray bands) originates from the synchrotron emission of relativistic electrons, whereas the high energy peak (GeV-TeV gamma-ray bands) can be explained by the leptonic models (SSC or EC), lepto-hadronic (p$\gamma$) and hadronic (pp) models (e.g., Ghisellini et al. 2010; Dermer et al. 2012; Zhang et al. 2012; Bottcher et al. 2013; Yan et al. 2014; Yan \& Zhang 2015; Ding et al. 2017; Zheng et al. 2017; Chen 2018; Xue et al. 2019a, 2019b; Gao et al. 2019). For the most powerful FSRQs ($L_{\rm \gamma} > 10^{48} {\rm erg~s^{-1}}$), there is often a third peak from the contribution of accretion disk (Ghisellini et al. 2017). According to the different synchrotron peak frequency ($\nu^s_{\rm{peak}}$), blazars can be classified as low-synchrotron-peaked blazars (LSPs; $\nu^s_{\rm{peak}}<10^{14}$ Hz), intermediate-synchrotron-peaked blazars (ISPs; $10^{14}<\nu^s_{\rm{peak}}<10^{15}$ or $10^{15.3}$ Hz) and high-synchrotron-peaked blazars (HSPs; $\nu^s_{\rm{peak}}>10^{15}$ or $10^{15.3}$ Hz; Abdo et al. 2010a; Fan et al. 2016). 

The high amplitude of flux variations and high luminosity are the distinctive properties of blazars (Wagner \& Witzel 1995). Most of flux variations in blazars are unpredictable, and only very few have relatively reliable period, such as OJ 287 (Valtonen et al. 2008; Zheng et al. 2008). The mechanisms of producing these unpredictable variations are under debate. The proposed mechanisms include injection, acceleration and cooling of particles, with possible intervention of shock waves or turbulence. A geometric interpretation is also proposed to explain the flux variations (Raiteri et al. 2017). For the convenience of reference, in view of different timescales and amplitudes of flux variations (e.g., Wagner \& Witzel 1995; Bai et al. 1999; Fan 2005; Fan et al. 2009, 2014, 2017; Gupta et al. 2008; Xiong et al. 2016, 2017), the flux variations of blazars are classified as three categories of intraday variability (IDV), short-term variability (STV) from days to weeks and long-term variability (LTV) from months to years. The flux variations of blazars in different timescales may have different origins. An effective pattern to investigate the origin of flux variations is to analyse the relationships between magnitude and color index (or flux and spectral variations). The relationships between magnitude and color index could represent a smooth transition from an accretion disk dominant color, to a mixed color from disk and jet, to a jet dominant color (see Fig. 4 from Isler et al. 2017). The flux variations of different blazars may correspond to different color behaviours depending on the contributions from jet and accretion disk components. Thus, it is possible to take flux variations of different blazars under a consolidated framework of color variations (see Isler et al. 2017 for more details). 

The bluer when brighter (BWB) chromatic trend was dominant for most of blazars, whereas the redder when brighter (RWB) trend was also found, especially for FSRQs (e.g., Gu et al. 2006; Bonning et al. 2012; Ikejiri et al. 2011). A RWB during the faint state and a slight BWB (or saturation trend) as the flux increases were found in some FSRQs (e.g., Raiteri et al. 2008, 2017; Fan et al. 2018). If the thermal component from accretion disk dominated the total radiation, accretion disk model could also produce the BWB trend (Liu et al. 2016; Gu \& Li 2013; Li \& Cao 2008). Isler et al. (2017) highlighted that 3C 279 was hard to reconcile with the simple RWB behaviour which is often associated with color variations from FSRQs. The analysis of Villata et al. (2004) revealed that the flux variations of the BL Lacertae (the prototype of BL Lac objects) could be interpreted in terms of two components: longer-term variations of mildly-chromatic events and a strong BWB of very fast (intraday) flares. Villata et al. (2002, 2004) suggested that the achromatic color trends were likely due to a change of the viewing angle on a ``convex'' spectrum for the BL Lacertae. This result was supported by Larionov et al. (2010). However, flux variations in some blazars could not be explained by the Doppler factor variations (e.g., Gaur et al. 2019). The above results indicate that the relationships between magnitude and color index in blazars are complex and related with the underlying physical mechanisms.

S5 0716+714 is an intermediate-synchrotron-peaked BL Lac object ($\nu^s_{\rm{peak}}=10^{14.66\pm0.02}{\rm Hz}$; Chen 2014). The position of optical band is almost equal to that of synchrotron peak. The source is known for its smooth non-thermal continua without emission lines (Danforth et al. 2013). Due to the lack of emission lines, the redshift is hard to be determined. Nilsson et al. (2008) claimed the redshift $z=0.31\pm0.08$ using the host galaxy as a standard candle, which was confirmed by the results from MAGIC Collaboration et al. (2019) and Danforth et al. (2013). The host galaxy has an $I$-band magnitude of 17.5$\pm$0.5 magnitudes (Nilsson et al. 2008). Because of its high power and lack of signs for ongoing accretion or surrounding gas, the source is an ideal candidate to explore the non thermal emission of jet (Vittorini et al. 2009). For the target, a strong BWB chromatic trend was found in the long-term, short-term, and intraday timescales (Poon et al. 2009; Wu et al. 2007; Dai et al. 2015; Dai et al. 2013; Yuan et al. 2017), whereas a mild BWB trend or no BWB trend was also claimed (Raiteri et al. 2003; Ghisellini et al. 1997; Chandra et al. 2011; Hu et al. 2014; Stalin et al. 2006; Agarwal et al. 2016; Hong et al. 2017; Zhang et al. 2018). Villata et al. (2000) discovered a FWB trend during fast variations and a steepening long-term trend of the spectral index using the Whole Earth Blazar Telescope (WEBT). The 2014 WEBT campaign targeting S5 0716+714 was organized to monitor the source. A BWB trend had been found in the source light curve, but no tight correlation between the source flux and color could be established (Bhatta et al. 2016). Villata et al. (2008) reported on the multifrequency behaviour of the source in 2007. Their results presented that the SEDs had the BWB behaviour (especially in the UV band).

\begin{figure}[hb!]
\begin{center}
\includegraphics[angle=0,scale=0.42]{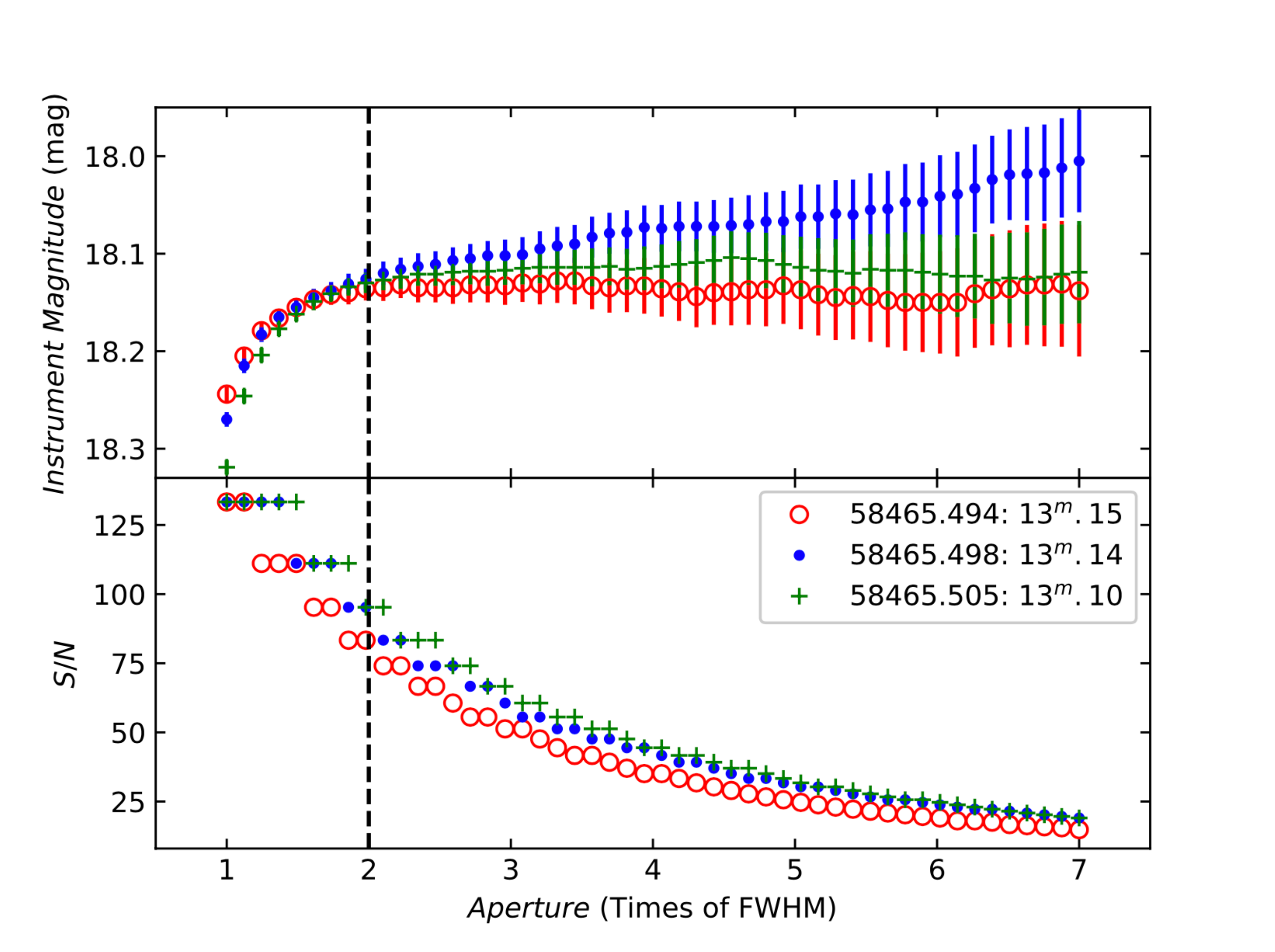}
		\caption{The aperture radii versus instrument magnitude and signal to noise ratio (S/N). The S/N is approximately equal to $\frac{1}{\rm error}$ in which error is measured by IRAF. The different colors stand for different MJDs. The vertical dashed line is $2\times{\rm FWHM}$.\label{fig1}}
\end{center}
\end{figure}

In order to explore the optical spectral behaviours (i.e., color behaviours) and flux variability in the long-term timescales, we continuously monitored the source in the optical bands from Nov. 10, 2017 to Jun. 06, 2019, and analysed the long-term spectral behaviours and flux variability. This paper is structured as follows. Section 2 gives the observations and data reduction. The results are reported in Section 3. Section 4 presents discussion. The conclusions are summarized in Section 5. The $\Lambda$CDM (concordance) cosmology with $H_{0}=67.74~{\rm km~s^{-1}~Mpc^{-1}}$, $\Omega_{\rm m}=0.309$, $\Omega_{\rm \Lambda}=0.691$ is adopted throughout this work (Planck Collaboration et al. 2016).

\section{OBSERVATIONS AND DATA REDUCTION}

Optical observations were carried out using the 60 cm BOOTES-4 auto-telescope, which was in a global network of robotic observatories. The fourth station of the BOOTE, BOOTES-4, is located at the Lijiang Observatory of the Yunnan Observatories of China (altitude of 3193 m). It is dedicated to observe optical emission from gamma-ray bursts (GRBs) and transients in the universe. Blazars are observed in the rest of time. The auto-telescope has been equipped with SDSS $u$, $g$, $r$, $i$ and $z$ filters since Mar. 2012. It has a field of view of $10\times10~{\rm arcmin^2}$ and an EMCCD camera with the pixel size of $13\times13~{\rm \mu m}$ and with the chip size of $1024\times1024$ pixels. The pixel scale corresponds to 0.56 arcsec/pixel. The readout noise and gain are 16 electrons and 1.75 electrons/ADU, respectively. 

The optical $g$, $r$ and $i$ bands are exposed in turn. The time intervals between two adjacent bands (e.g., $r$ and $i$, $g$ and $r$) at the same observational sequence ($gri$-bands) range from 22 seconds to 130 seconds with an average value of 46 seconds, depending on weather conditions. The total number of observations is 201 nights including 26973 data points (8991 points per band). The observational log in the $r$ band is given in Table 1. The logs from other bands are almost same as that of Table 1. Table 1 shows that the sampling intervals range from 1 day to 7 days with an average value of 1.5 days during the first observed season (MJD=58067 to 58253), and from 1 day to 21 days with an average value of 3 days during the second observed season (MJD=58433 to 58640). The average time spans and time resolutions are 3.4 hours and 2.9 minutes per night, respectively.

After correcting flat field and bias, we use the DAOPHOT task of IRAF\footnote{IRAF is distributed by the National Optical Astronomy Observatories, which are operated by the Association of Universities for Research in Astronomy, Inc., under cooperative agreement with the National Science Foundation.} to implement aperture photometry. The aperture radius of 2$\times$FWHM, the inner radius of 5$\times$FWHM and width of 2$\times$FWHM in the sky annulus are set up. Ideally, one comparison star is enough for differential photometry. However, it is hard to find a comparison star with the same color and magnitude as the target source in actual observation. The errors from variations of comparison stars and accidental probability of only one comparison star should be considered. Therefore more than one comparison star is often adopted. The standard stars 5 and 6 in the finding chart\footnote{https://www.lsw.uni-heidelberg.de/projects/extragalactic/charts/0716+714.html} are chosen as comparison stars to get the target source magnitude because their magnitudes and locations are more close to the target source and they have the apparent magnitude in the $I$ band. 

In order to determine the optimal aperture radius, we select 50 different aperture radii from $1\times{\rm FWHM}$ to $7\times{\rm FWHM}$ on MJD=58465.494, 58465.498, 58465.505 when the magnitudes are near average magnitude of the whole data. The aperture radii versus instrument magnitude and signal to noise ratio (S/N) are given in Fig. 1. The Fig. 1 shows that contrary to the trend of instrument magnitude, the overall trend in the S/N is decreasing with increasing aperture radii. The instrument magnitudes are roughly invariable using apertures from $1.6\times{\rm FWHM}$ to $2.5\times{\rm FWHM}$. At the same time, in light of the changes of brightness on different days, the aperture radius of 2$\times$FWHM is our best choice.

The $UBVRI$-magnitudes of comparison stars 5 and 6 are taken from Ghisellini et al. (1997) and Villata et al. (1998). The empirical color transformations between SDSS photometry and Johnson-Cousins UBVRI for stars are taken to calculate the $gri$-magnitudes of the two comparison stars (Jordi et al. 2006; star 5: $13^{\rm m}.804$, $13^{\rm m}.367$, $13^{\rm m}.261$; star 6: $13^{\rm m}.89$, $13^{\rm m}.447$, $13^{\rm m}.37$). The rms errors of photometry on a certain night are calculated as (Bai et al. 1998; Fan et al. 2014; Xiong et al. 2017):
\begin{equation}
\sigma_{\rm rms}=\sqrt{\sum \frac{(m_i-\overline{m})^2}{N-1}},~~~~i=1,2,3,...,N,
\end{equation}
where $m_i = (m_{\rm 5}-m_{\rm 6})_{i}$ is the differential magnitudes of stars 5 and 6, $\overline{m}=\overline{m_{5}-m_{6}}$ is the averaged differential magnitudes, and $N$ is the number of the observations over the night. In addition, the Poisson errors from the target source and the comparison stars measured by IRAF should be considered (Liu et al. 2019; $err_{\rm iraf}=\sqrt{err_{\rm target}^2+{err_{\rm star5}^2+err_{\rm star6}^2}}$). The final errors are calculated as $\sigma_{\mathrm{err}}=\sqrt{\sigma_{\rm rms}^2+err_{\rm iraf}^2}$. The table of data is available in Table 2.

\section{RESULTS}

The $gri$-magnitudes are corrected for Galactic extinction. The extinction values are obtained from the NASA/IPAC Extragalactic Database\footnote{http://ned.ipac.caltech.edu} (NED; $A_{\rm i}=0.053, A_{\rm r}=0.071, A_{\rm g}=0.102$). The long-term $r$ band light curves show that the target source brightens from $14^{\rm m}.16$ to $12^{\rm m}.29$ together with five prominent sub-flares in the first observed season, and first becomes fainter from $12^{\rm m}.89$ to $14^{\rm m}.76$ and then brightens from $14^{\rm m}.76$ to $12^{\rm m}.94$ with seven prominent sub-flares in the second observed season (Fig. 2). The average values are $13.29^{\rm m}\pm0.01$ and $13.60^{\rm m}\pm0.01$ respectively for the two observed seasons, and the overall flux variability amplitudes (Heidt \& Wagner 1996) both about $1^{\rm m}.87$.
 
\begin{figure}
\begin{center}
\includegraphics[angle=0,scale=0.42]{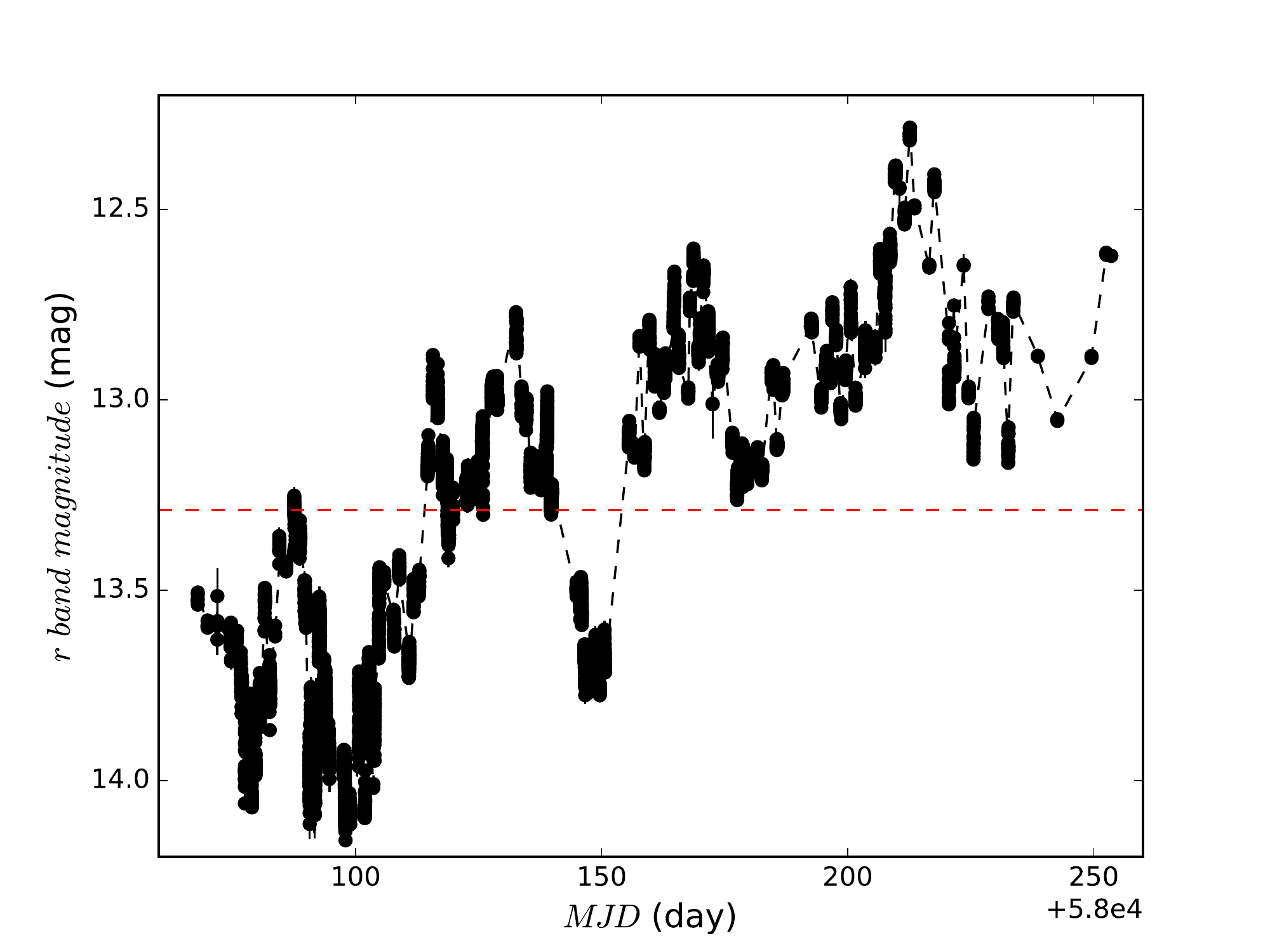}
\includegraphics[angle=0,scale=0.42]{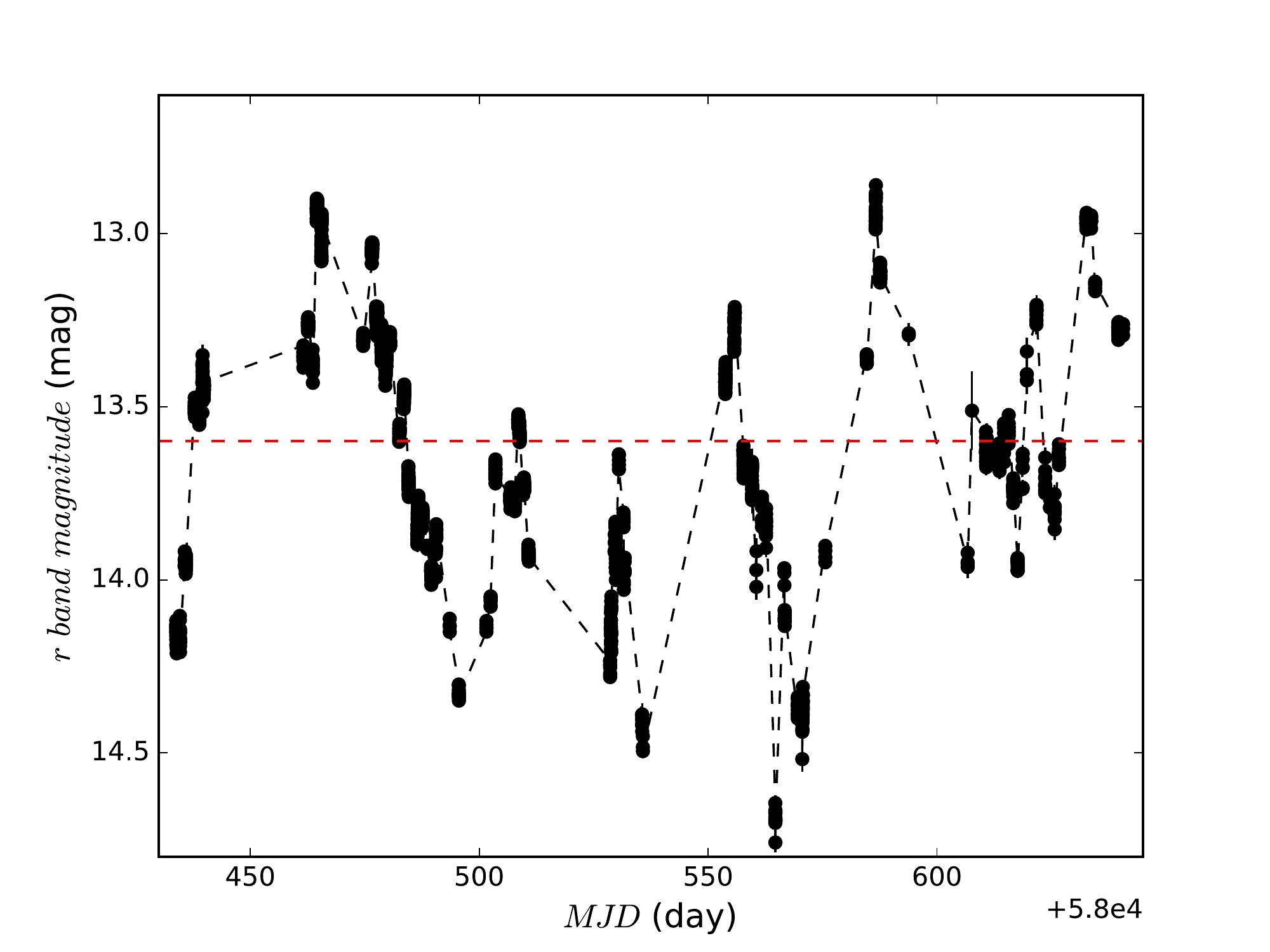}
\caption{Long-term $r$ band light curves in the first and the second observed seasons. The red dashed lines are average values for each panel.\label{fig2}}
\end{center}
\end{figure}

\begin{figure}
\begin{center}
\includegraphics[angle=0,scale=0.24]{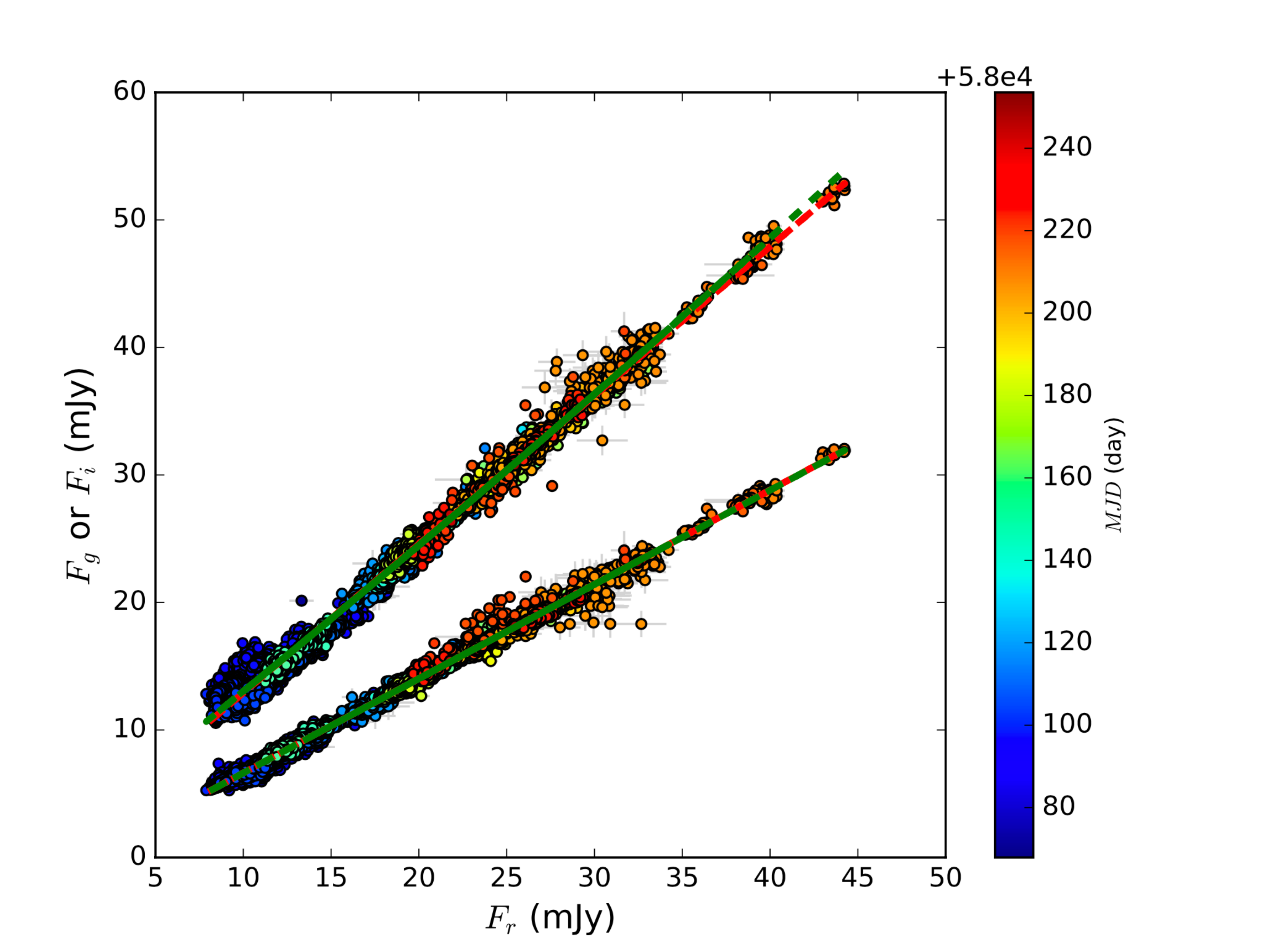}
\includegraphics[angle=0,scale=0.24]{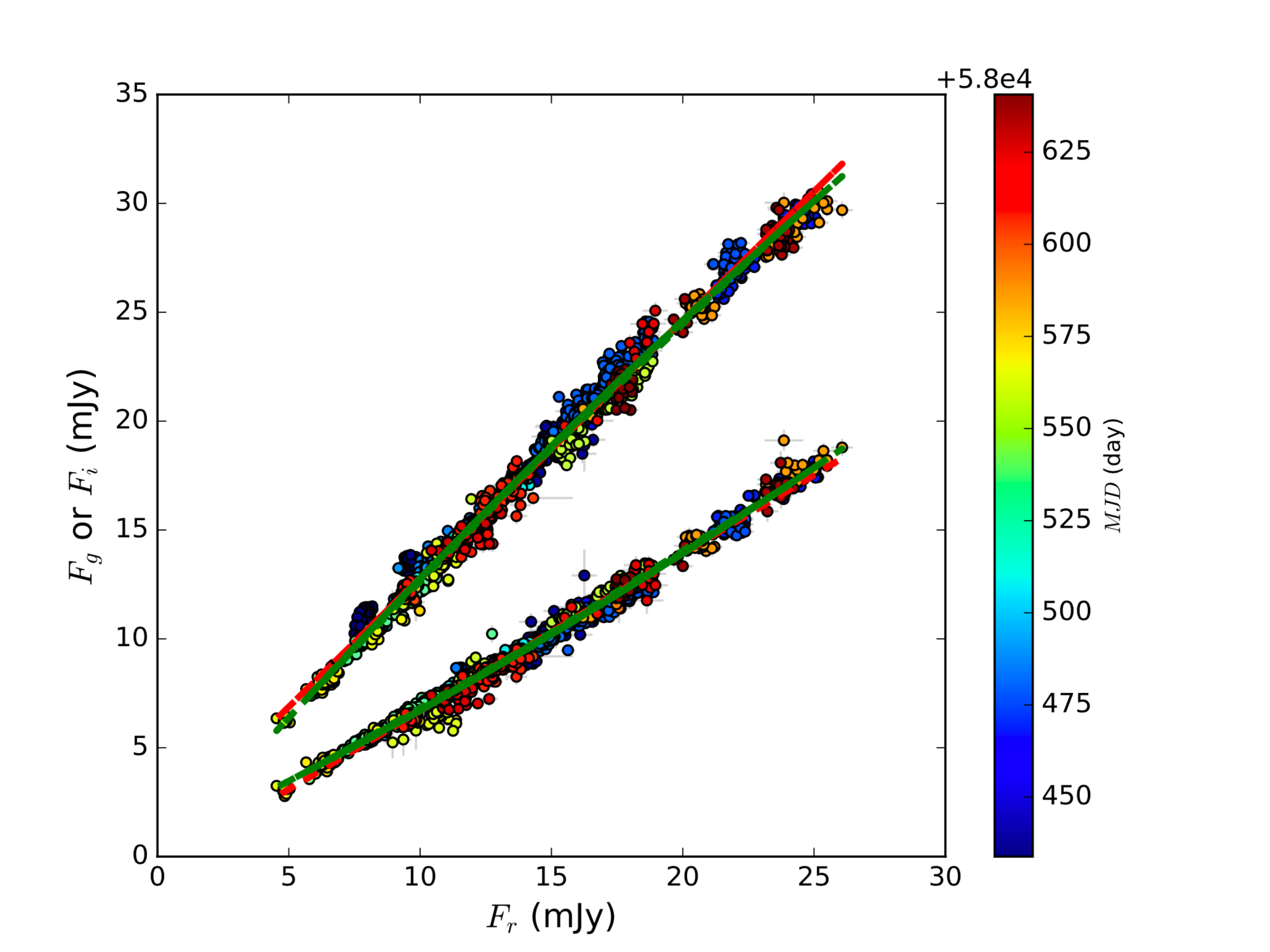}
\caption{The $r$ band flux versus $g$ or $i$ band flux. The color bars are the MJD. The green solid line is the results from linear fitting of least square, and the red dashed line from the second order polynomial fitting of least square. The two lines almost overlap each other because the coefficient of quadratic term is small. The left panel indicates that this target is observed in the first observation season and the right panel indicates that this target is observed in the second observation season. \label{fig3}}
\end{center}
\end{figure}

We find that the zero-points in the $g$, $r$ and $i$ bands are close to the zero-point in the AB system by referring to photometric flux calibration\footnote{http://classic.sdss.org/dr4/algorithms/fluxcal.html} of SDSS filters. The AB system is defined such that every filter has a zero-point flux density of 3631 Jy. Thus the $gri$-magnitudes are converted into fluxes by using $F=3631\times10^{-0.4*mag}\times10^3$ mJy. The least square polynomial fitting is used to analyse the relationships between $r$ band flux and $g$ (or $i$) band flux (Fig. 3). The Fig. 3 displays that the relationships between $r$ band flux and $g$ (or $i$) band flux can be best fitted by a linear model, rather than the second order polynomials. Even if all data observed from 58067 to 58640 are considered, the above results still remain. Such results are different from that of concave regression from Larionov et al. (2010), and demonstrate the optical spectrum in the $gri$-bands expressed likely by a power law ($F\propto\nu^{-\alpha}$). For a power law intrinsic spectrum, if the Doppler factor $\delta$ and $\alpha$ are uniform among the $gri$-bands, the variations of Doppler factor can not make the spectral index (or color) change since the frequencies $\nu=\delta\nu'$ and the flux $F_{\nu}(\nu)=\delta^{2+\alpha}F'_{\nu'}(\nu)$ for a continuous jet (Urry \& Padovani 1995). 

The logarithm of power law spectrum ($F_{\nu}\propto\nu^{-\alpha}$) can be written as ${\rm log}F_{\nu}=-\alpha{\rm log}\nu+C$ in which $C$ is a constant. If we have observational quasi-simultaneous fluxes from different frequencies, the slope ($-\alpha$) of linear fitting can be obtained. The linear fitting of weighted least square\footnote{https://www.statsmodels.org/stable/index.html} (WLS) is adopted to fit the quasi-simultaneous fluxes in the $gri$-bands ($g$ band: $6.2\times10^{14}$ Hz; $r$ band: $4.8\times10^{14}$ Hz; $i$ band: $3.9\times10^{14}$ Hz). The fitting error of slope is taken as the spectral index error. Fig. 4 shows the correlations between optical spectral index $\alpha_{\rm gri}$ and $r$ band flux $F_{\rm r}$ in the two observed seasons. The correlations can not be described by a single flatter when brighter (FWB) trend or steeper when brighter (SWB) trend, and vary with time. The results of WLS fitting get a strong FWB trend at a low flux state and then a weak FWB trend at a higher flux state in the first observed season (the correlation coefficient $r=0.69$, the chance probability $P<10^{-4}$ and the slope $k=0.072$ at a low flux state; $r=0.27$, $P<10^{-4}$ and $k=0.004$ at a higher flux state), and a weak FWB trend at a low flux state and then a strong FWB trend at a higher flux state in the second observed season ($r=0.19$, $P<10^{-4}$ and $k=0.009$ at a low flux state; $r=0.6$, $P<10^{-4}$ and $k=0.02$ at a higher flux state). The divided value is equal to $F_{\rm r}$=15 mJy. From the left panel in Fig. 4, there are different trends on the left and right sides of the divided value. The divided value is approximately equal to the average magnitude of the entire sampling data. Below the divided value, the source is considered at a low flux state and above the divided value, the source is considered at a higher flux state. We should note that the break flux (15 mJy) is estimated via visually identifying the patterns before and after the value, which could bring into uncertainty. However, the estimated break flux is unlikely to have a marked impact on the above results because the above results still exist when any of break fluxes ranging from $\sim$13 mJy to $\sim$20 mJy are selected.

\begin{figure}
\begin{center}
\includegraphics[angle=0,scale=0.24]{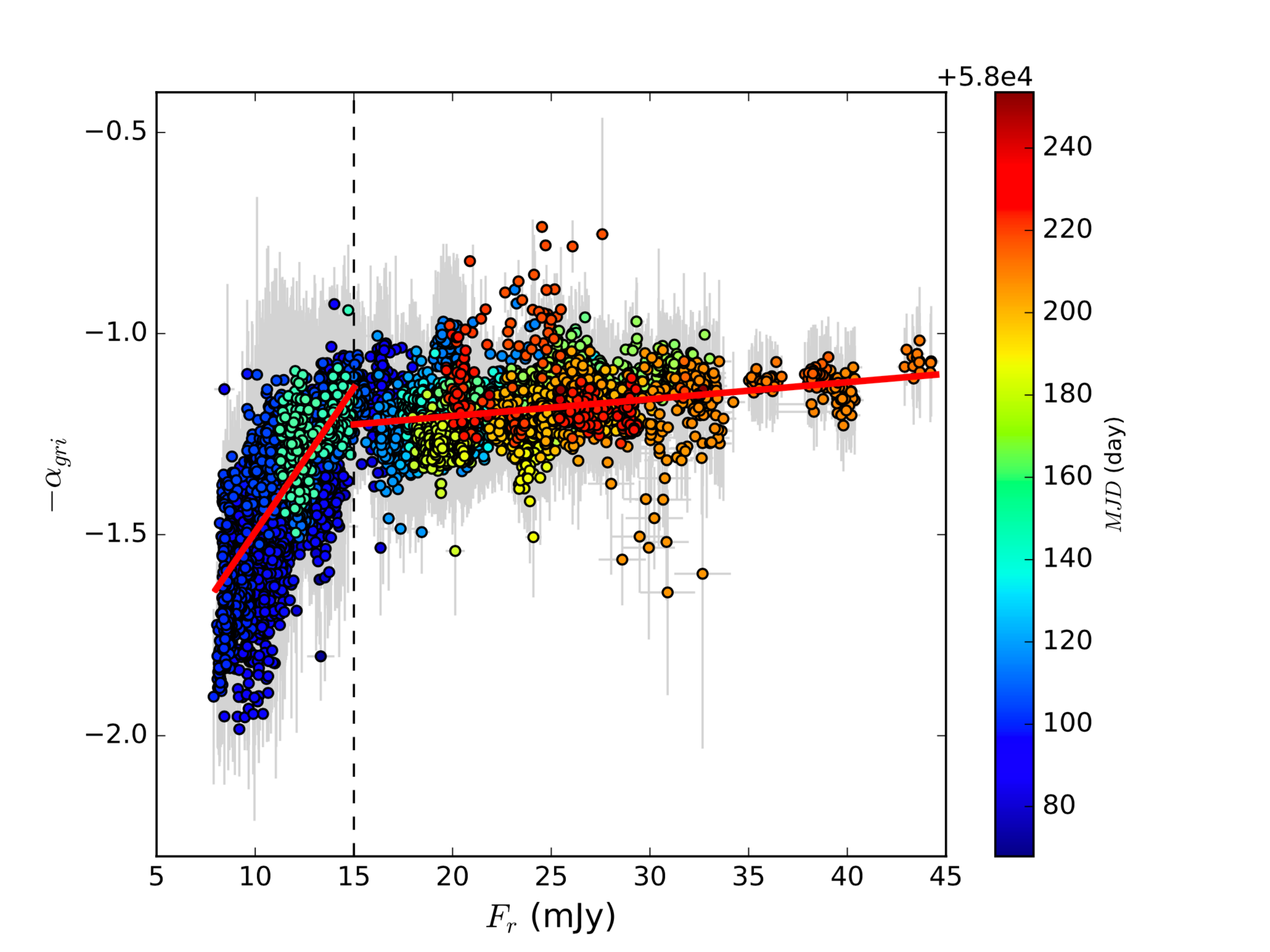}
\includegraphics[angle=0,scale=0.24]{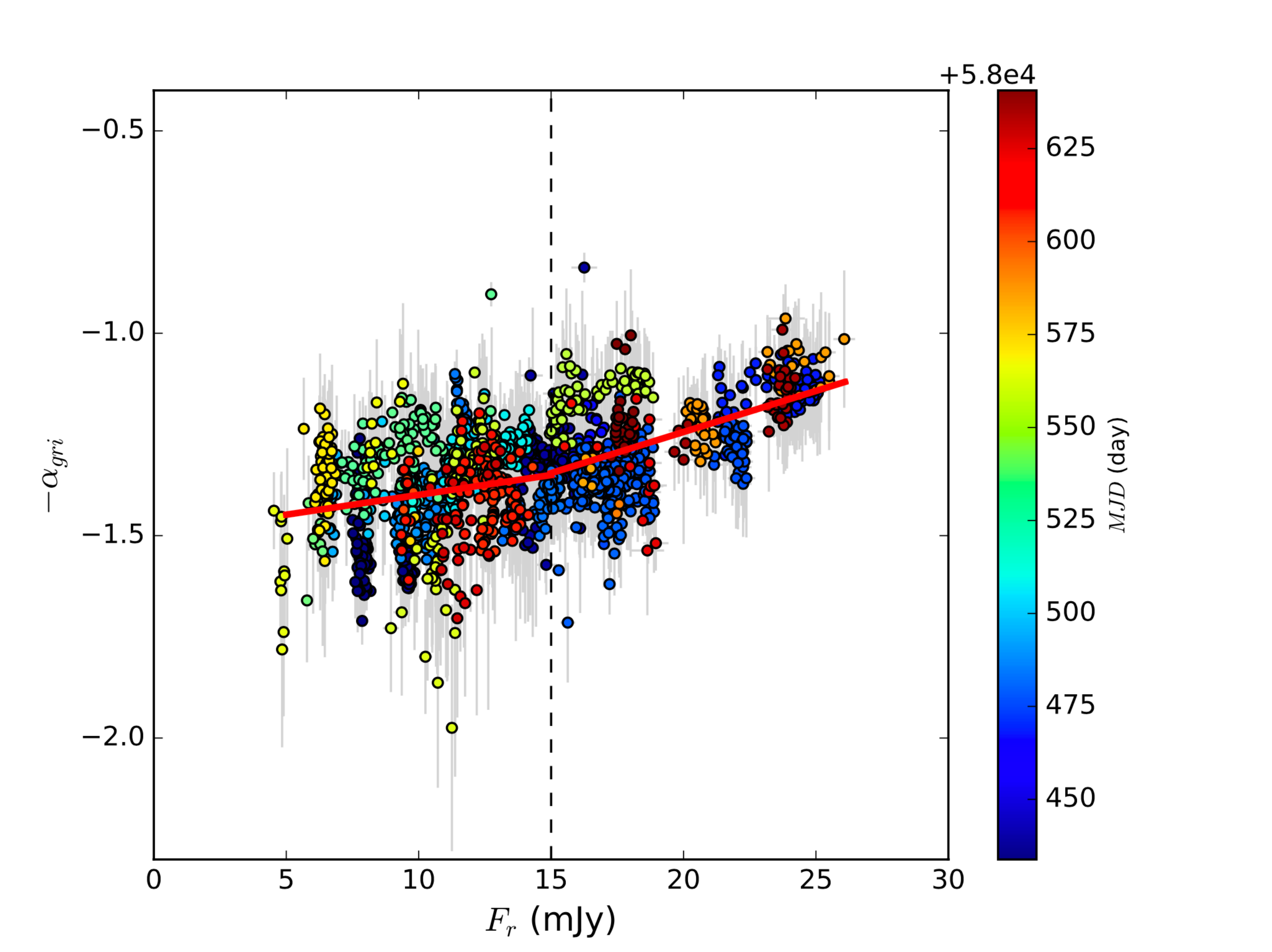}
\caption{The correlations between optical spectral index $\alpha_{\rm gri}$ and $r$ band flux $F_{\rm r}$. The red solid lines are the results of WLS fitting. The black dashed lines are divided lines ($F_{\rm r}$=15 mJy). The color bars are the MJD. The left panel indicates that this target is observed in the first observation season and the right panel indicates that this target is observed in the second observation season.\label{fig4}}
\end{center}
\end{figure}

\begin{figure}[htb!]
\begin{center}
\includegraphics[width=8.9cm,height=8.9cm]{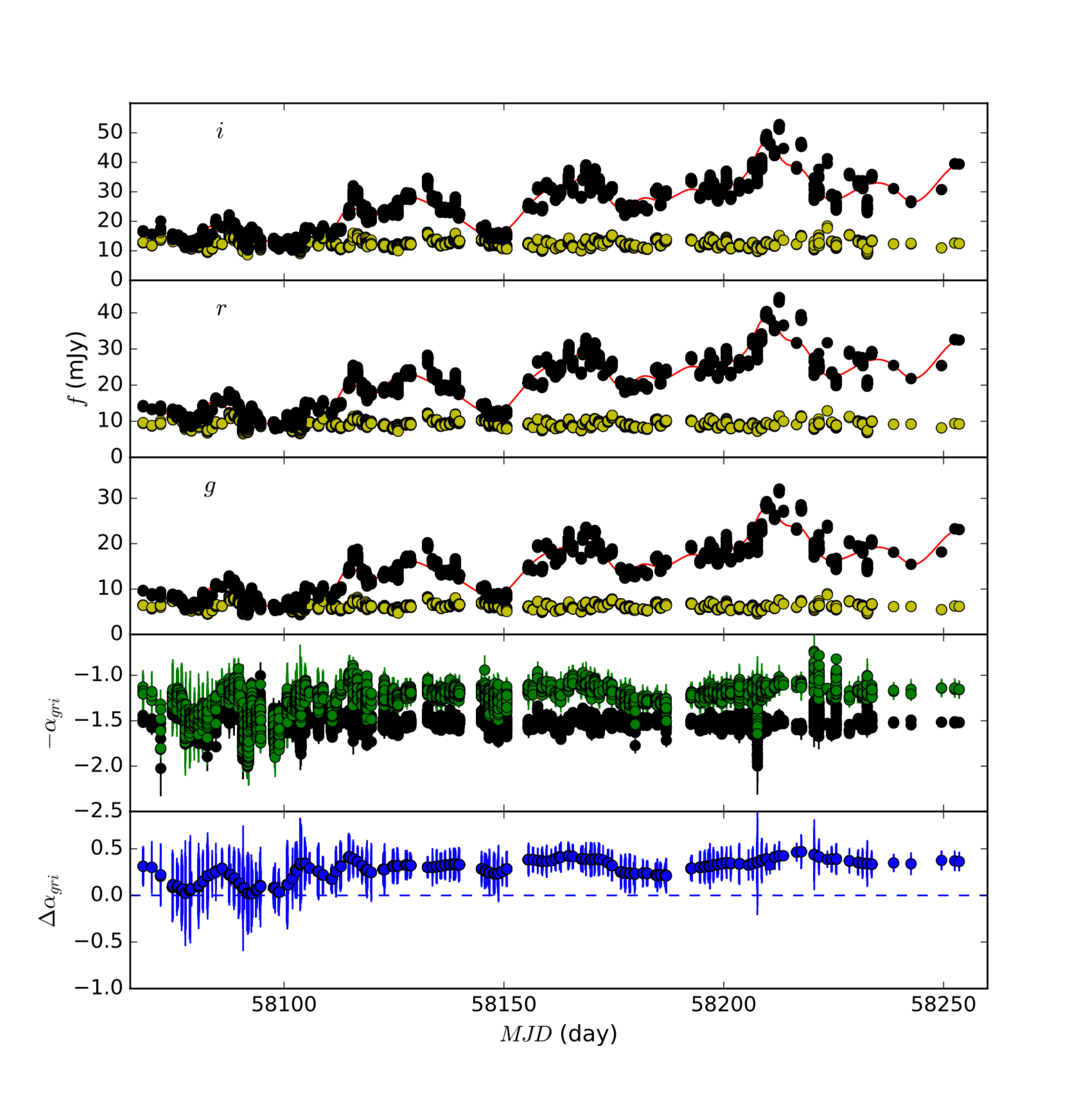}
\includegraphics[width=8.9cm,height=8.9cm]{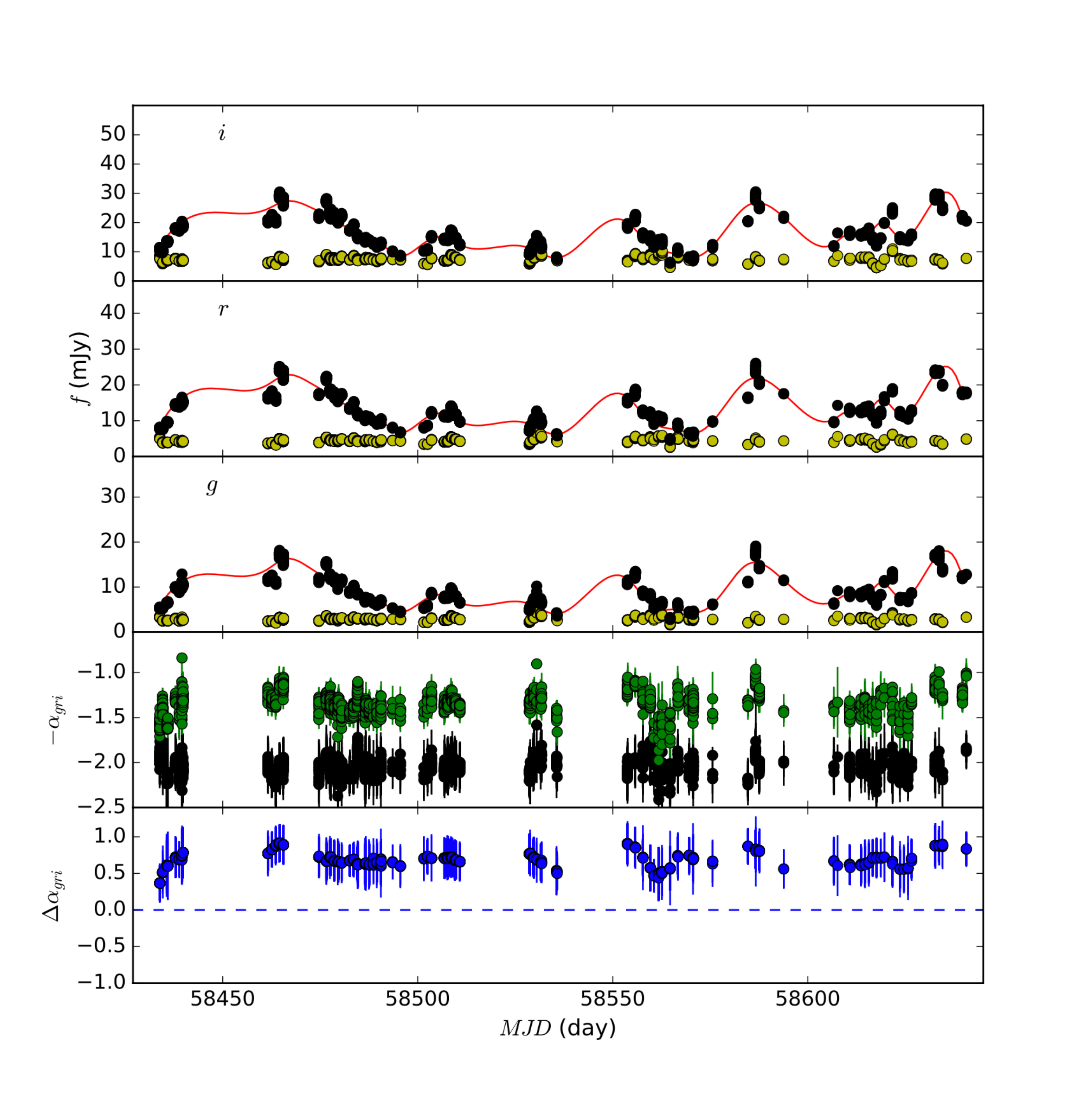}
\caption{The variations from light curve and spectral index for 4-days binning data. For the first three panels, the black circles and yellow circles are the original light curve and the light curve of removing the long-term trend, respectively; the red lines are the cubic spline interpolations of 4-day binning data. In the fourth panel, the green and black circles are the original spectral index and the spectral index of removing the long-term trend, respectively. In the last panel, the blue circles stand for the values of $\Delta \alpha_{\rm gri}$ and the blue dashed lines represent $\Delta \alpha_{\rm gri}=0$. These left panels indicate that this target is observed in the first observation season and these right panels indicate that this target is observed in the second observation season.\label{fig5}}
\end{center}
\end{figure}

\begin{figure}[htb!]
\begin{center}
\includegraphics[angle=0,scale=0.24]{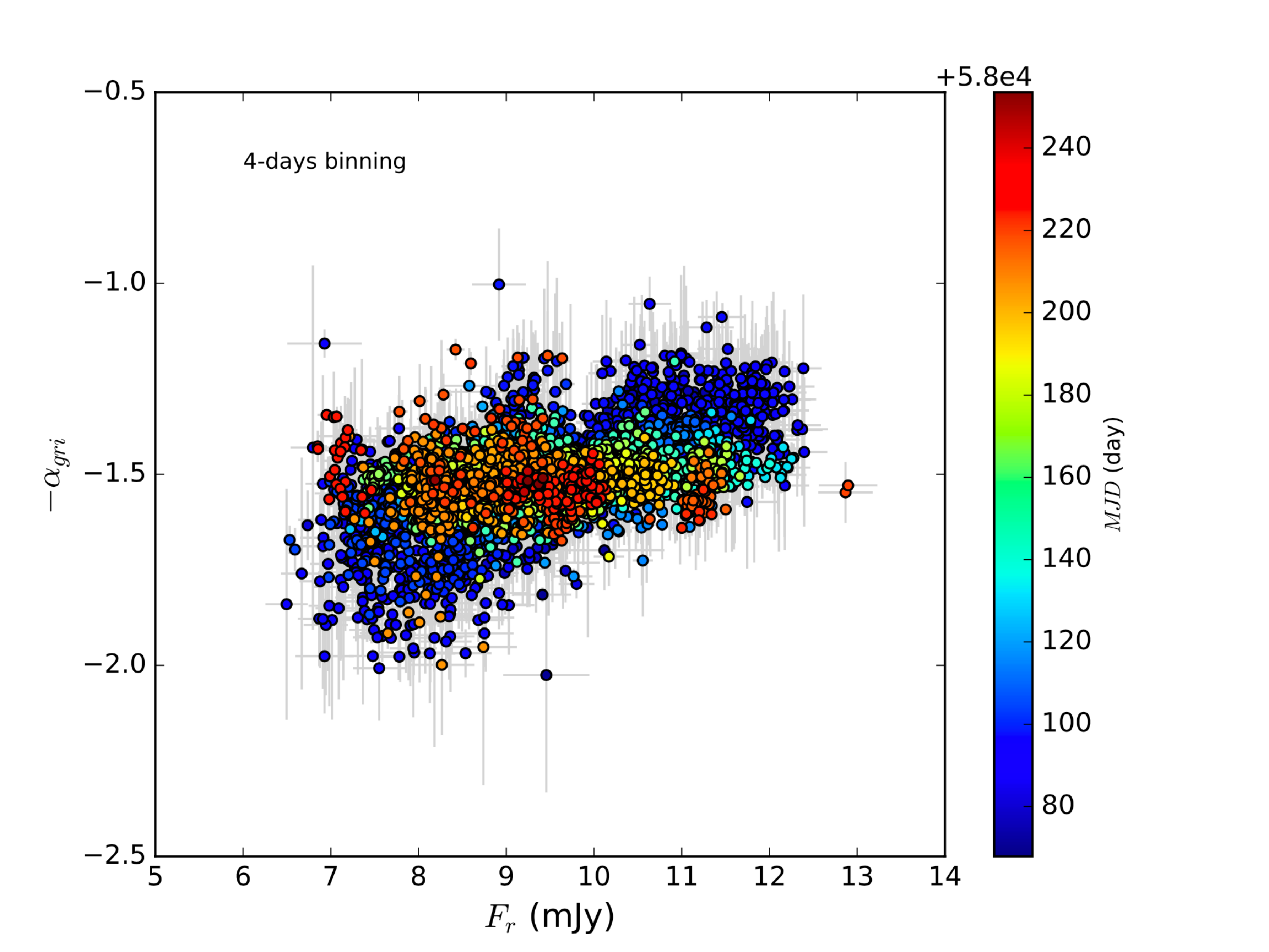}
\includegraphics[angle=0,scale=0.24]{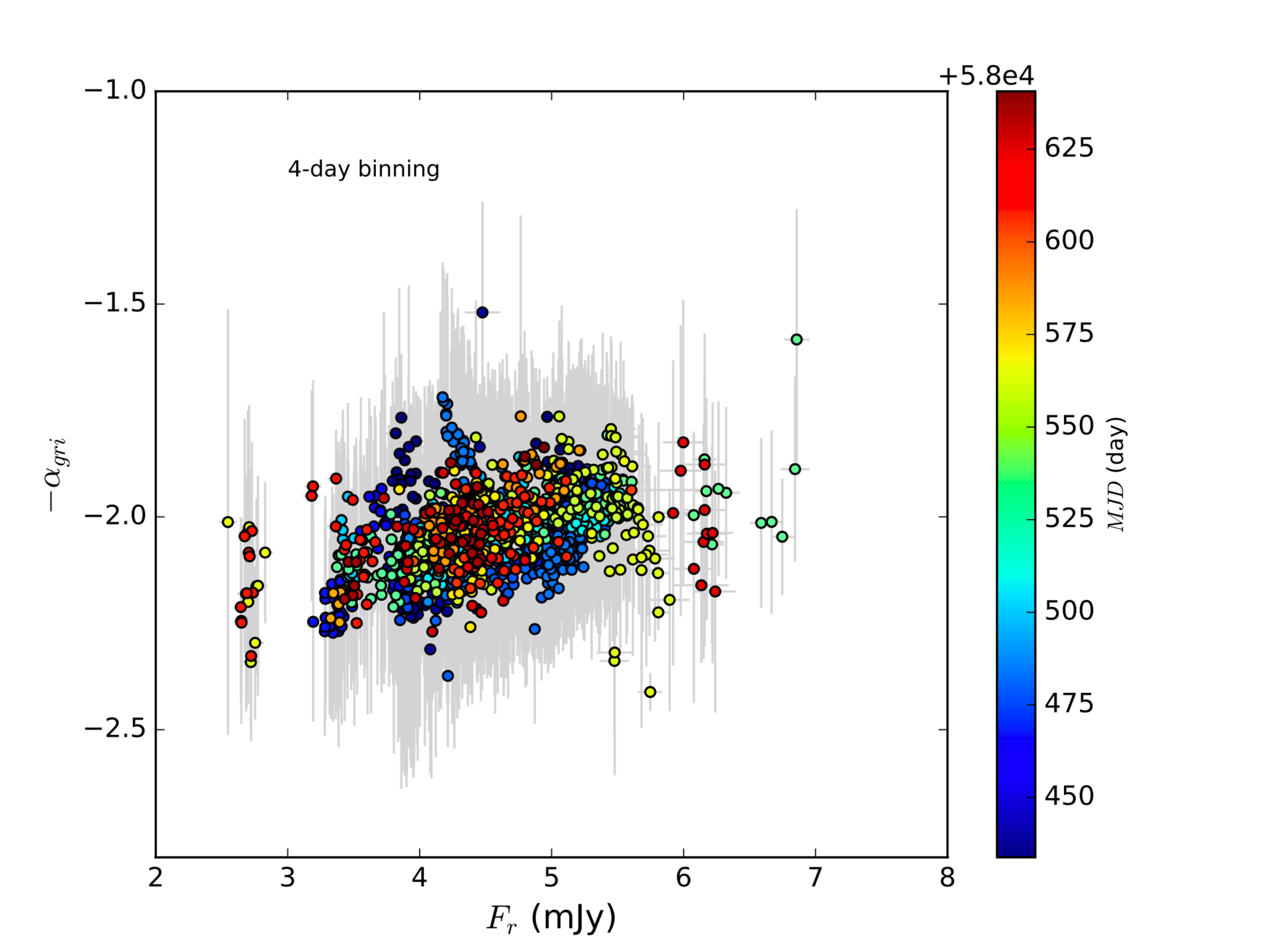}
\includegraphics[angle=0,scale=0.24]{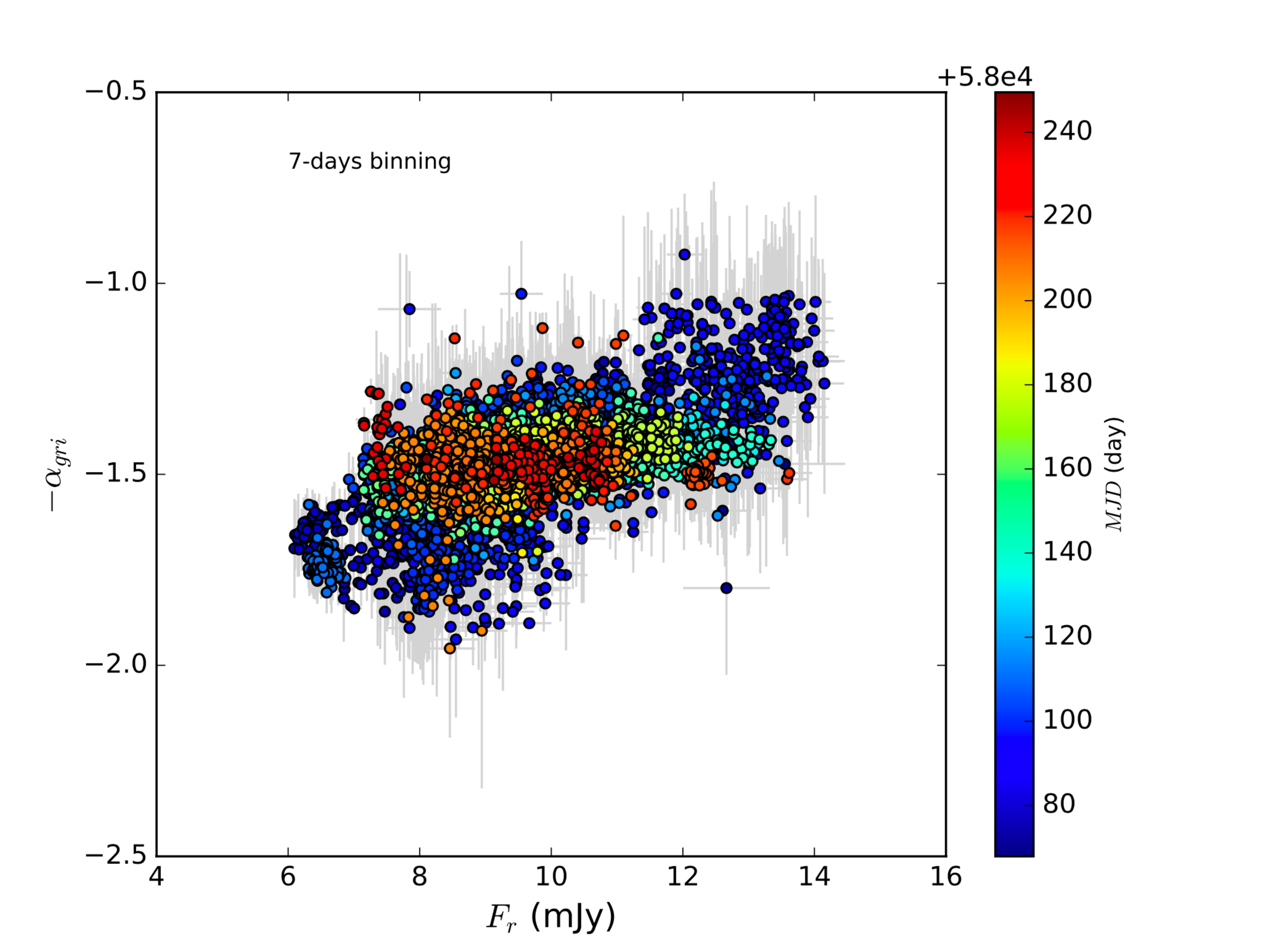}
\includegraphics[angle=0,scale=0.24]{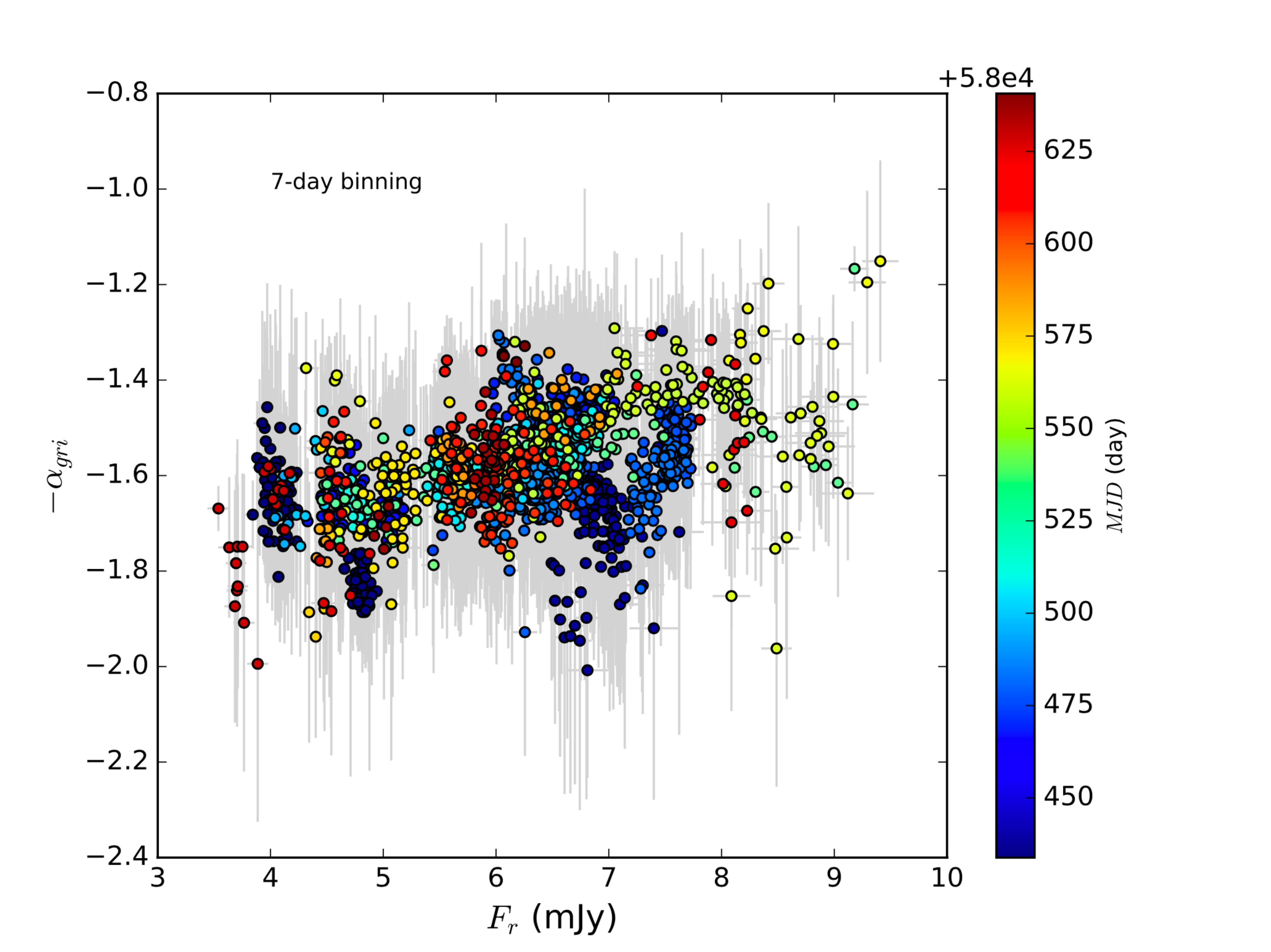}
\includegraphics[angle=0,scale=0.24]{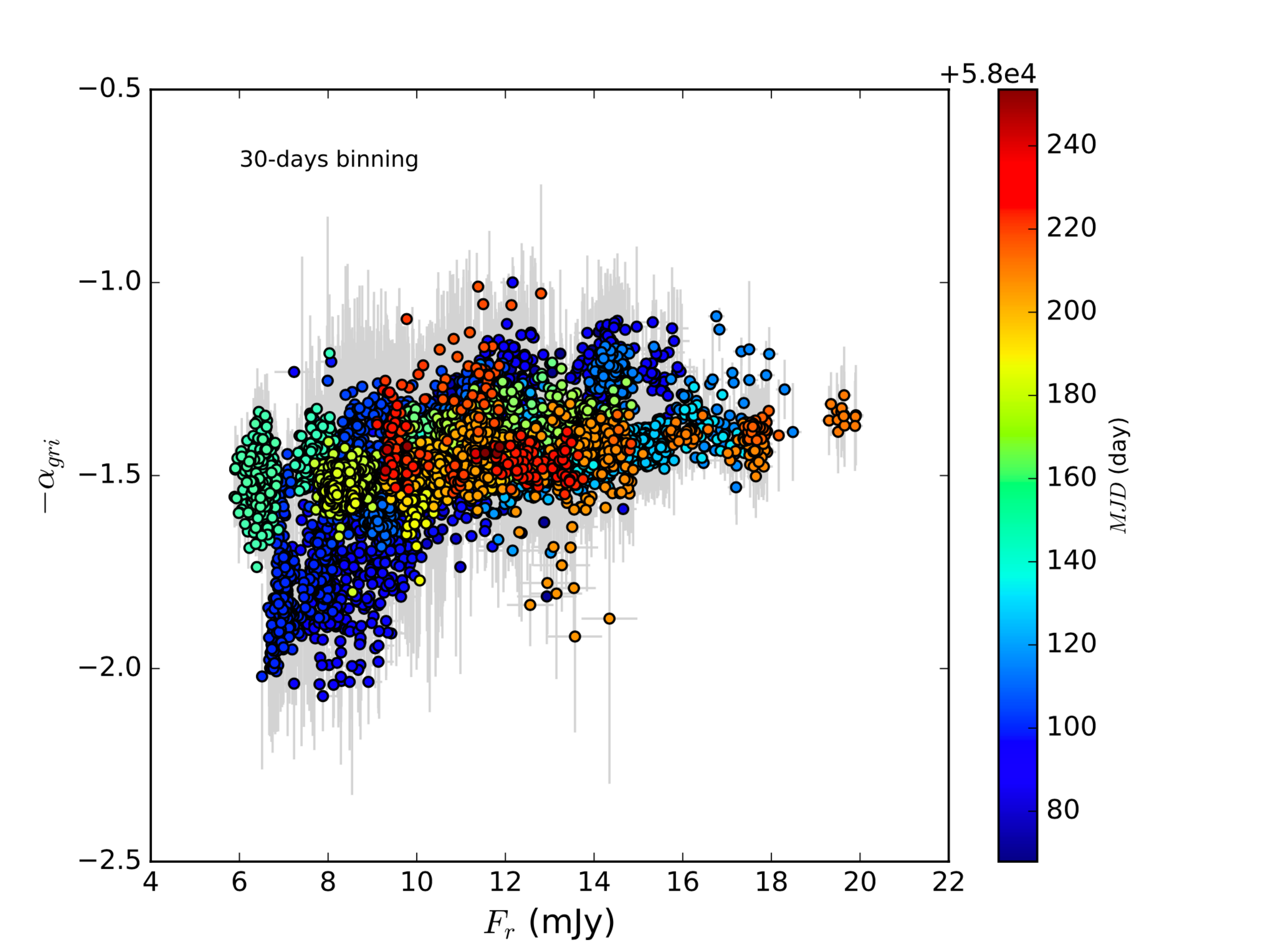}
\includegraphics[angle=0,scale=0.24]{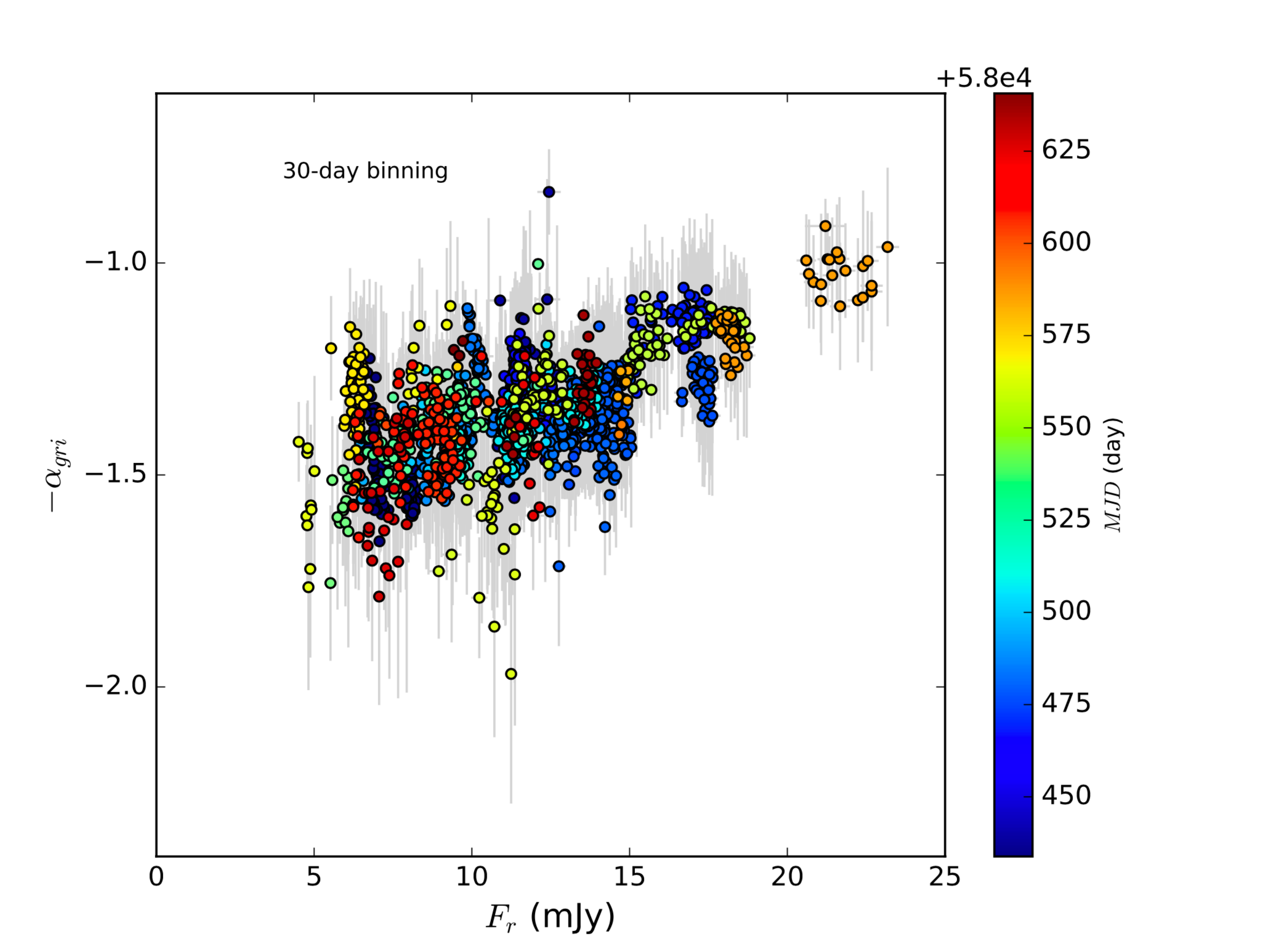}
\caption{The correlations between optical spectral index $\alpha_{\rm gri}$ and $r$ band flux $F_{\rm r}$ after the long-term trends of different binning data are removed. The left panel indicates that this target is observed in the first observation season and the right panel indicates that this target is observed in the second observation season.\label{fig6}}
\end{center}
\end{figure}

In order to explore whether the origins of the FWB trends are related with the variations of Doppler factor, following a procedure from Villata et al. (2004), we remove the long-term trend from the original light curve. The cubic spline interpolations\footnote{https://docs.scipy.org/doc/scipy/reference/generated/scipy.interpolate.UnivariateSpline.html\#scipy.interpolate.UnivariateSpline} of 4-day binning data are adopted to stand for the long-term trend. Each original flux is rescaled by dividing it by the ratio between the value of the spline at the considered time and its minimum value, $C_j(t)=[F_{\rm spl}(t)/F_{\rm min}]_j$ (where j represents band). The spectral index of removing the long-term trend ($\alpha{'}$) is estimated using the fluxes in $gri$-bands after the long-term variations are removed. The variations from light curve and spectral index can be found in Fig. 5. After the long-term trend of 4-days binning data is removed, most of the values of $\Delta \alpha_{\rm gri}$, which is equal to the absolute value of spectral index of removing the long-term trend subtracting the original spectral index ($\Delta \alpha_{\rm gri}=\mid\alpha{'}-\alpha\mid$), are not close to zero (Fig. 5). Although a moderate correlation across all data appears for the first observed season ($r=0.58$, $P<10^{-4}$), a strong FWB trend below the MJD=58120 ($r=0.69$, $P<10^{-4}$) and a weak FWB trend above the MJD=58120 ($r=0.19$, $P<10^{-4}$) still remain (Fig. 6). After the long-term trend of 4-days binning data is removed, a weak FWB trend above the MJD=58120 from Fig. 4 is not significantly improved to a middle/strong FWB trend in the first observed season. The weak FWB trend still remains for the results from 7-days binning data and 30-days binning data (Fig. 6). In the second observed season, after the long-term trends of 4-days and 7-days binning data are removed, the general FWB trends ($r=0.44$ and $r=0.38$, $P<10^{-4}$; the right panel of Fig. 6) become weak compared to the general FWB trend without removing the long-term trend ($r=0.6$, $P<10^{-4}$; the right panels of Fig. 4); after the long-term trend of 30-days binning data is removed, the general FWB trend ($r=0.61$, $P<10^{-4}$) is close to the general FWB trend without removing the long-term trend. Thus, after the long-term trend is removed, the general FWB trend is no significant improvement in the second observed season. Apart from the result of 30-days binning data in the second observed season, the results from 7-days binning data and 30-days binning data are consistent with the results of 4-days binning data, i.e., most of the values of $\Delta \alpha_{\rm gri}$ are not close to zero (compare Fig. 6 with Fig. 4). 

\begin{figure}[htb!]
\begin{center}
\includegraphics[angle=0,scale=0.29]{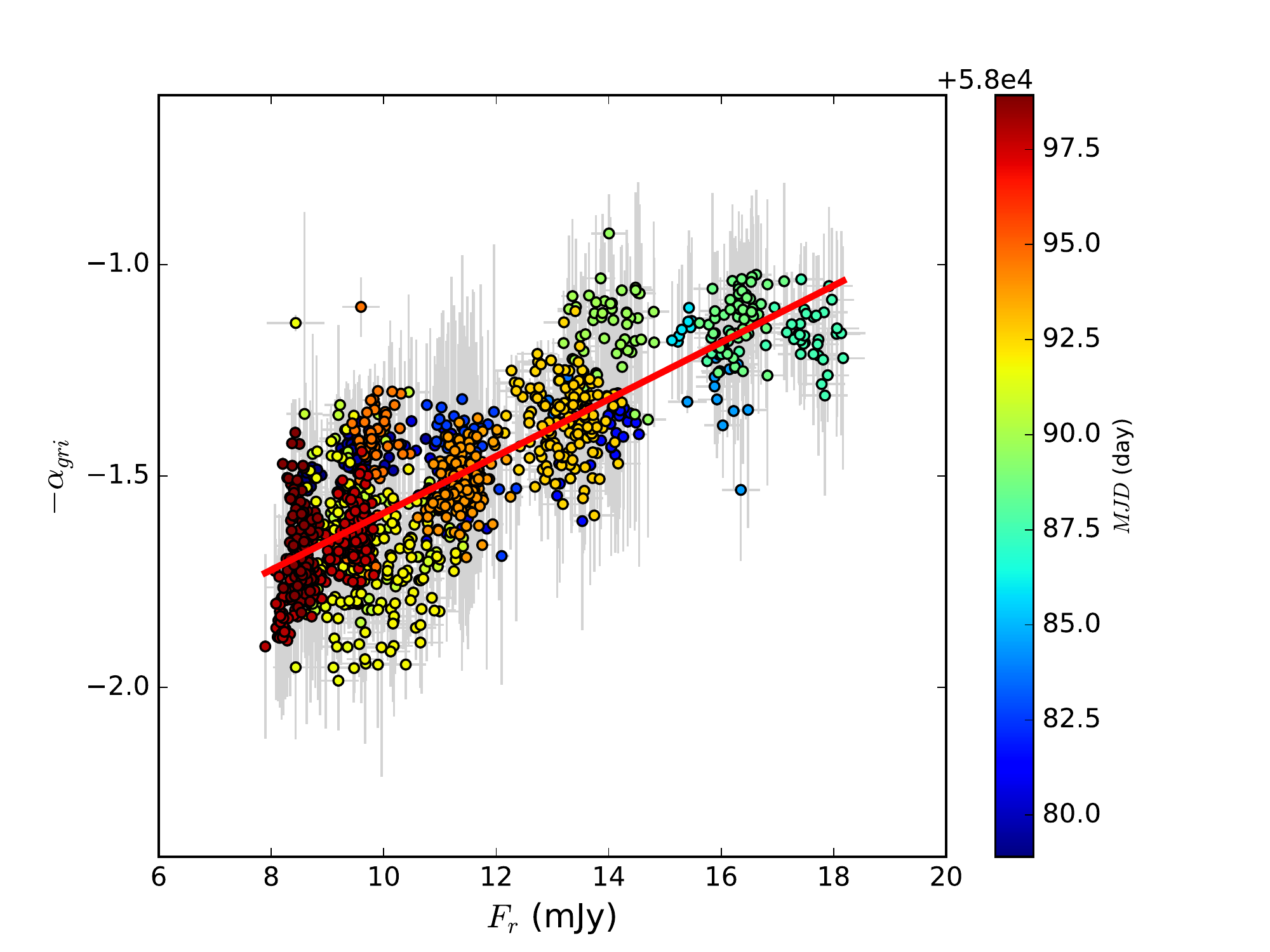}
\includegraphics[angle=0,scale=0.29]{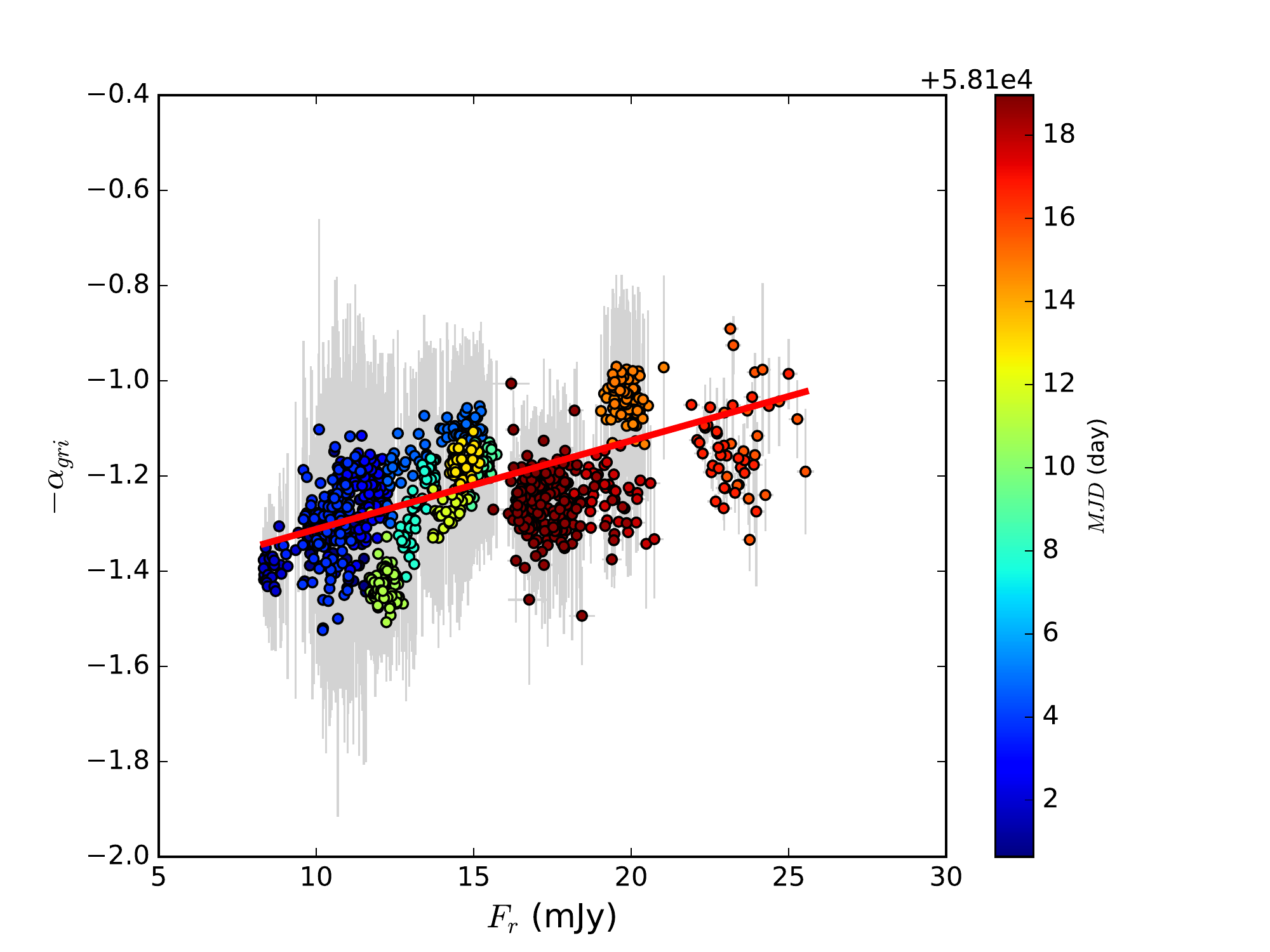}
\includegraphics[angle=0,scale=0.29]{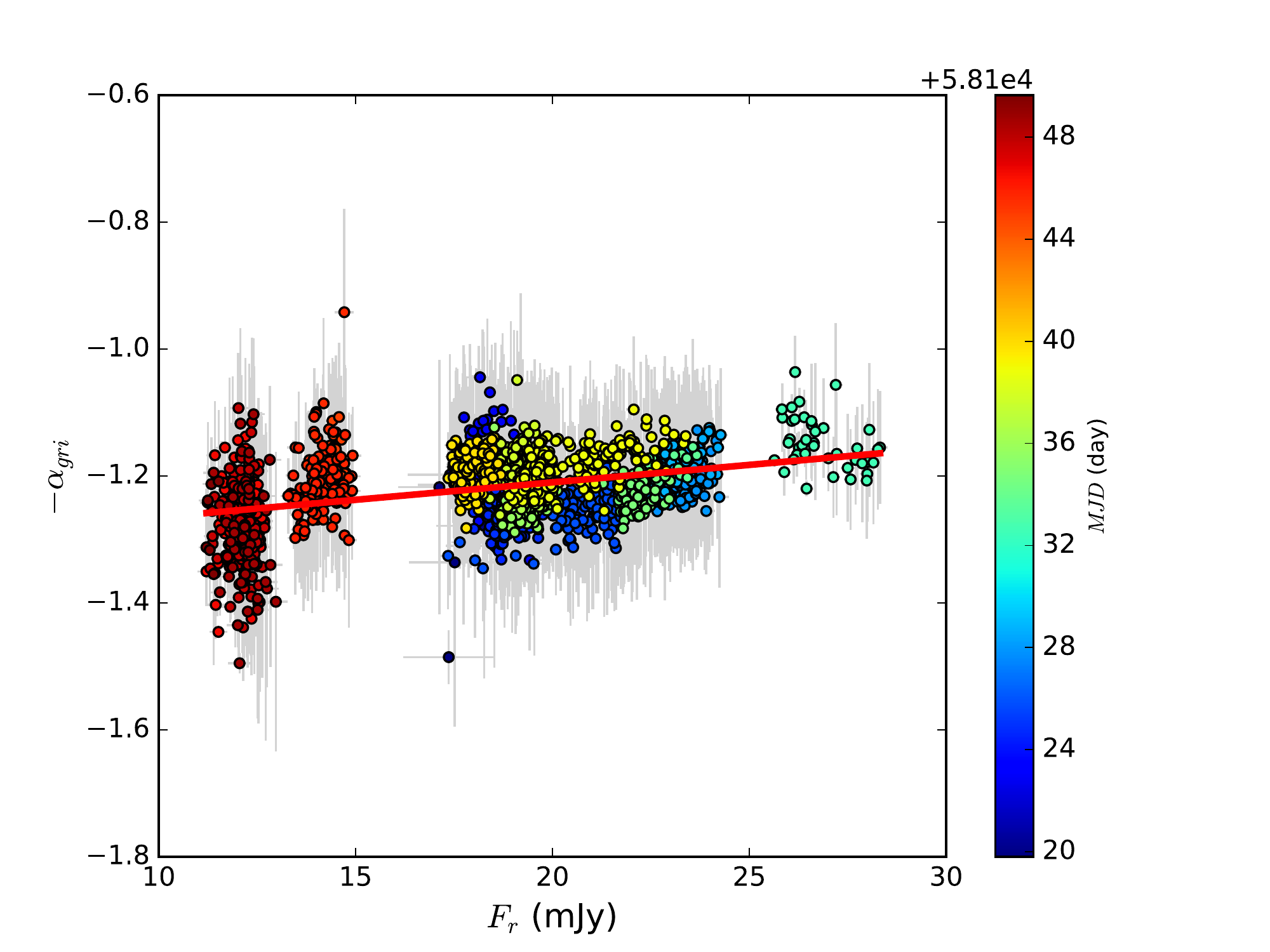}
\includegraphics[angle=0,scale=0.29]{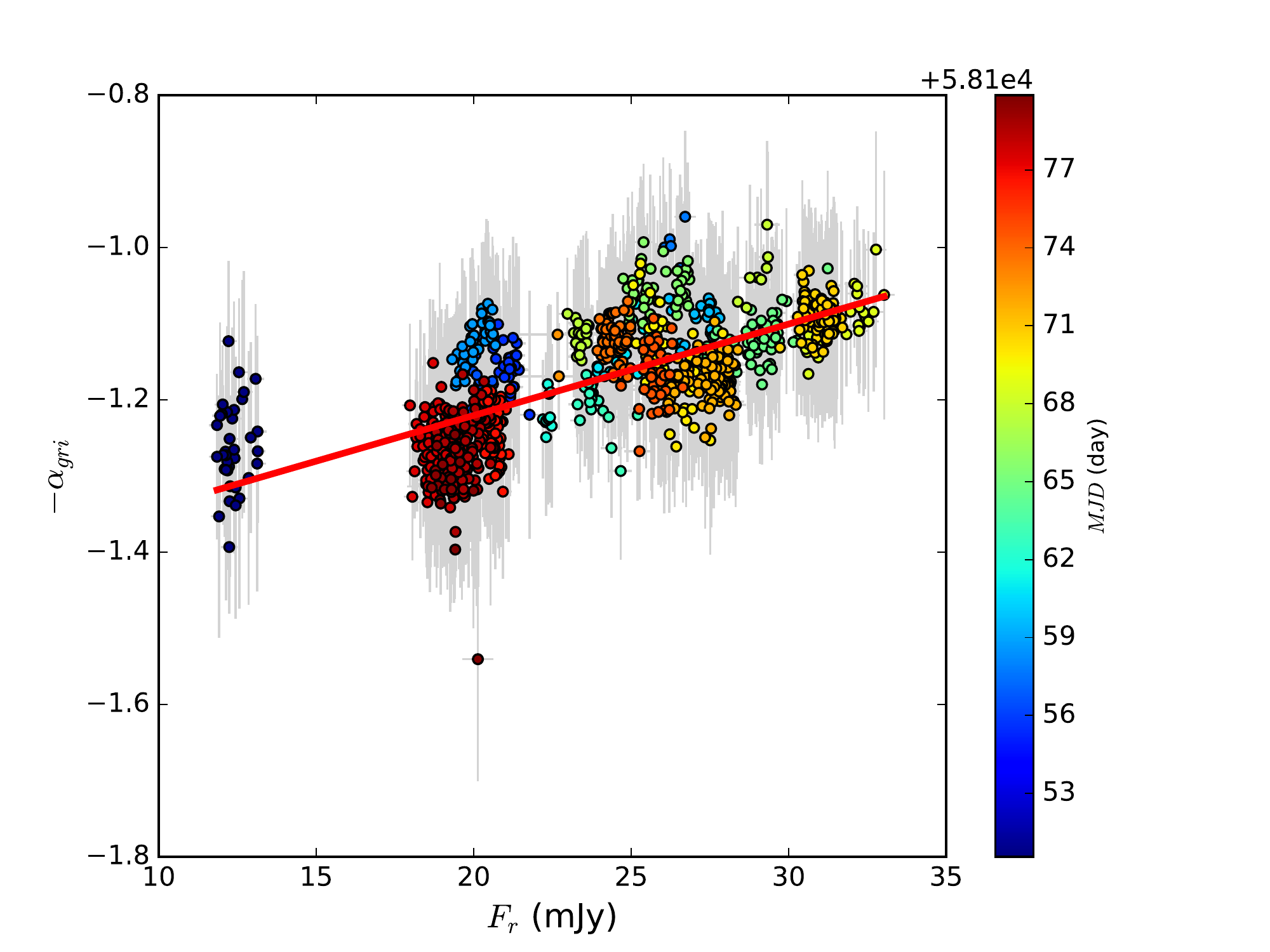}
\includegraphics[angle=0,scale=0.29]{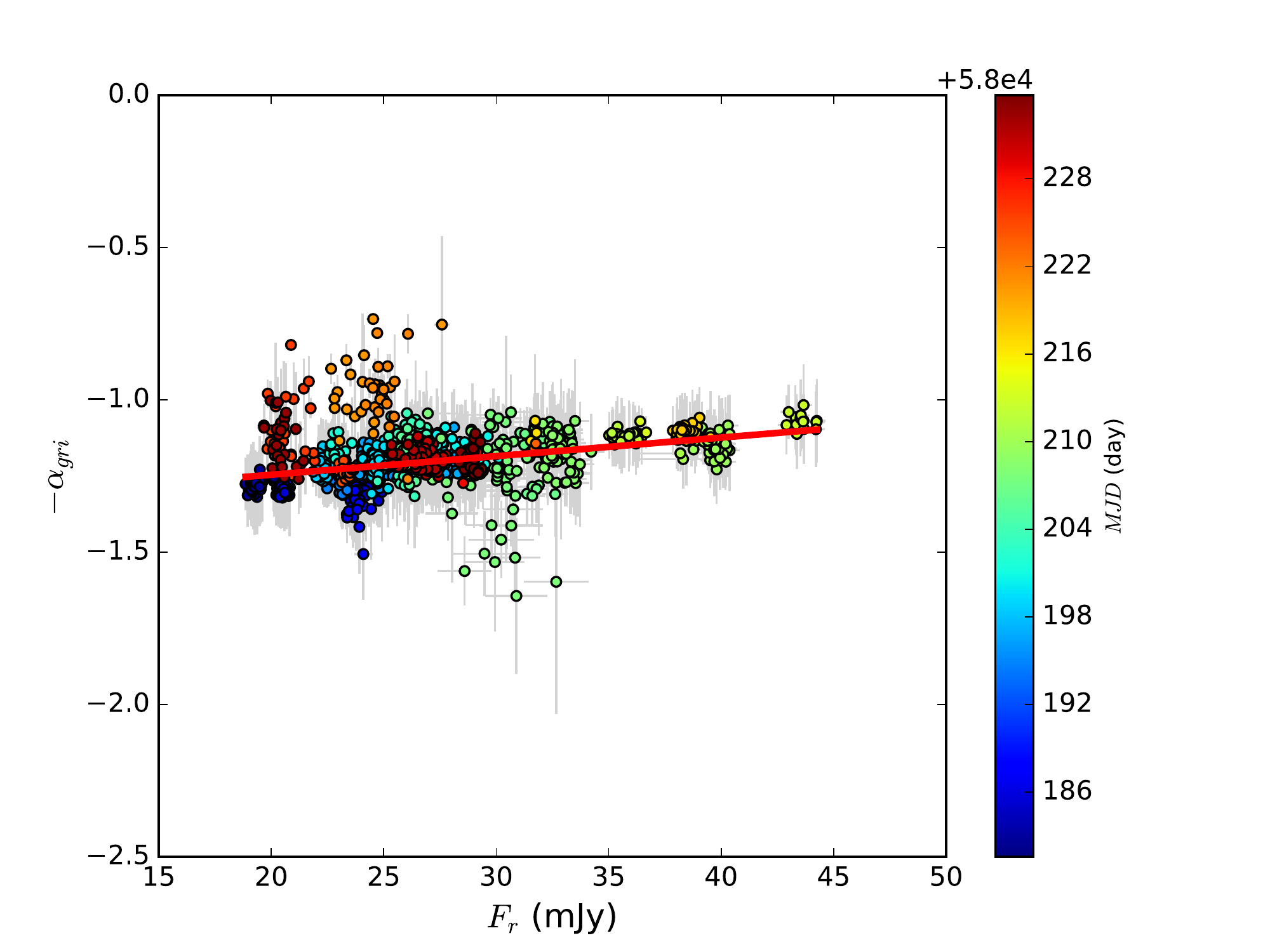}
\includegraphics[angle=0,scale=0.29]{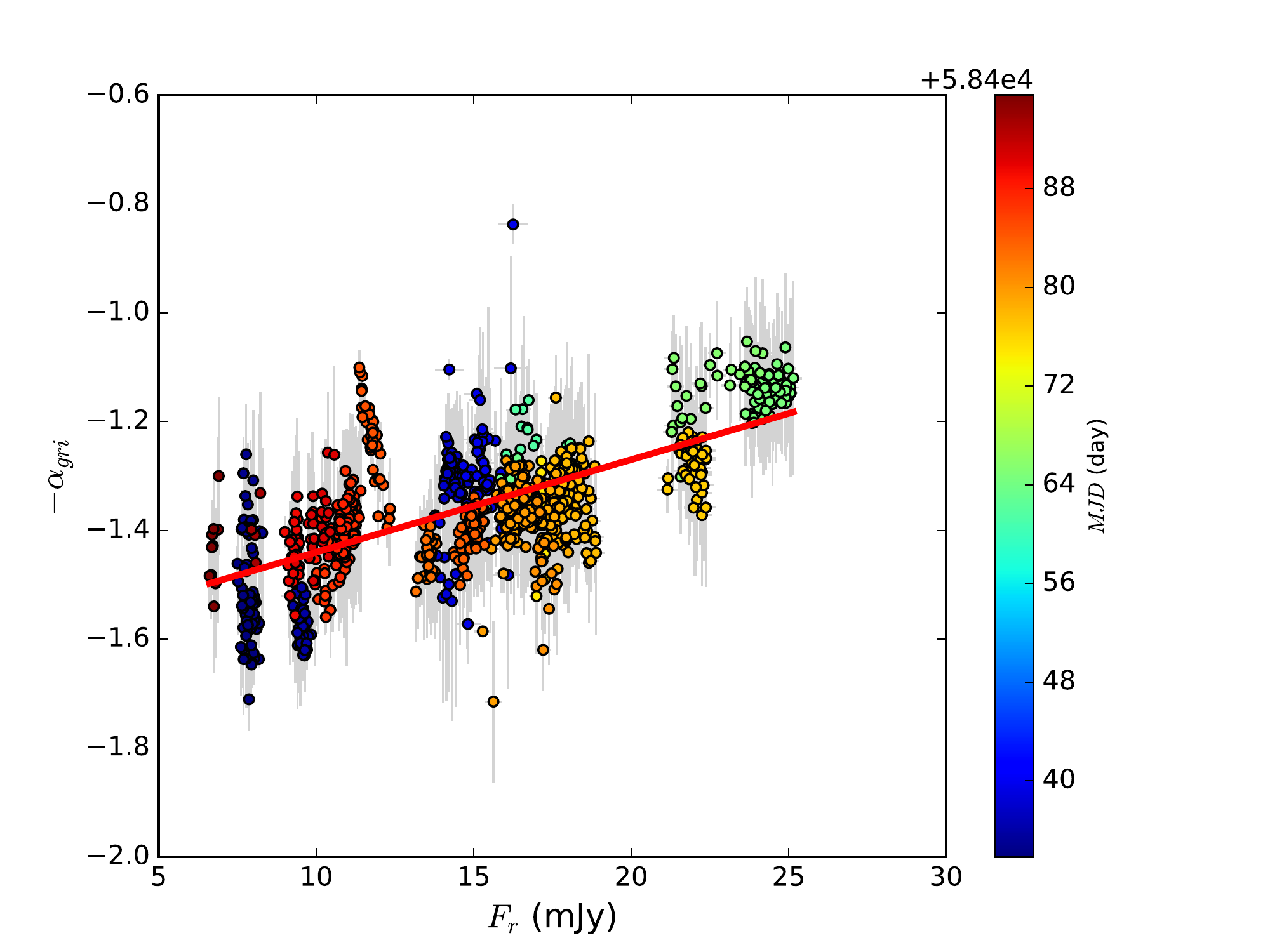}
\includegraphics[angle=0,scale=0.29]{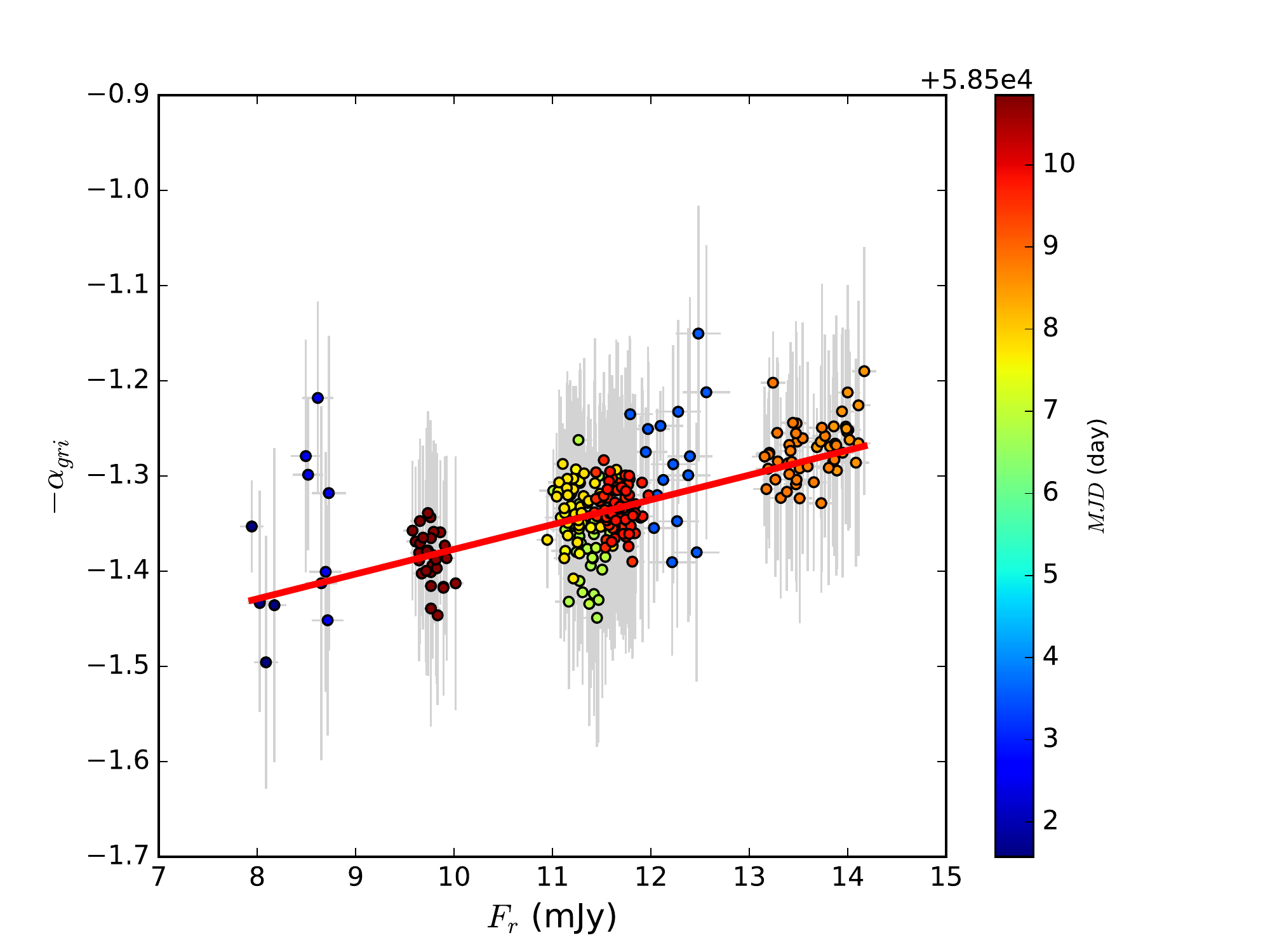}
\includegraphics[angle=0,scale=0.29]{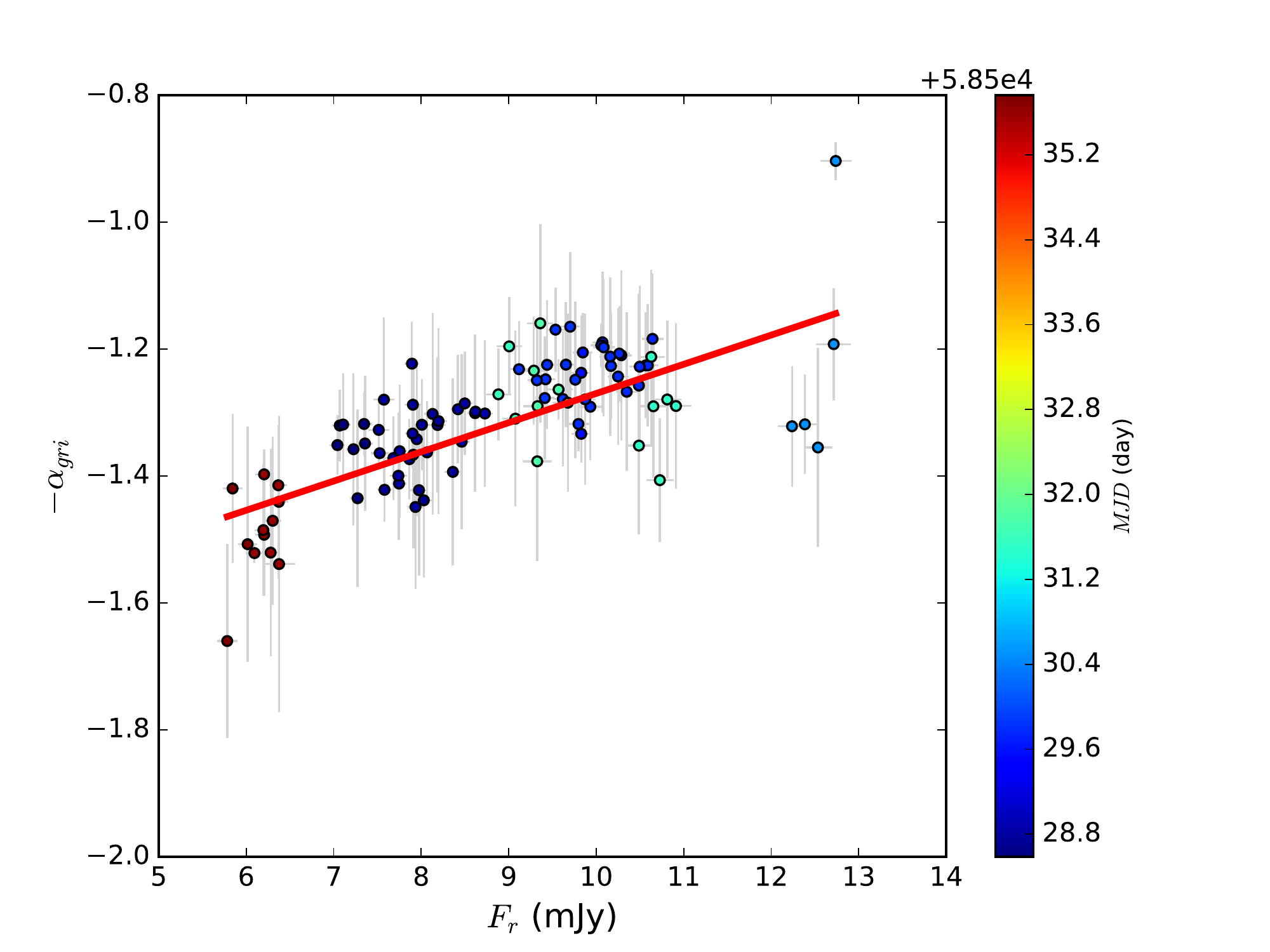}
\includegraphics[angle=0,scale=0.29]{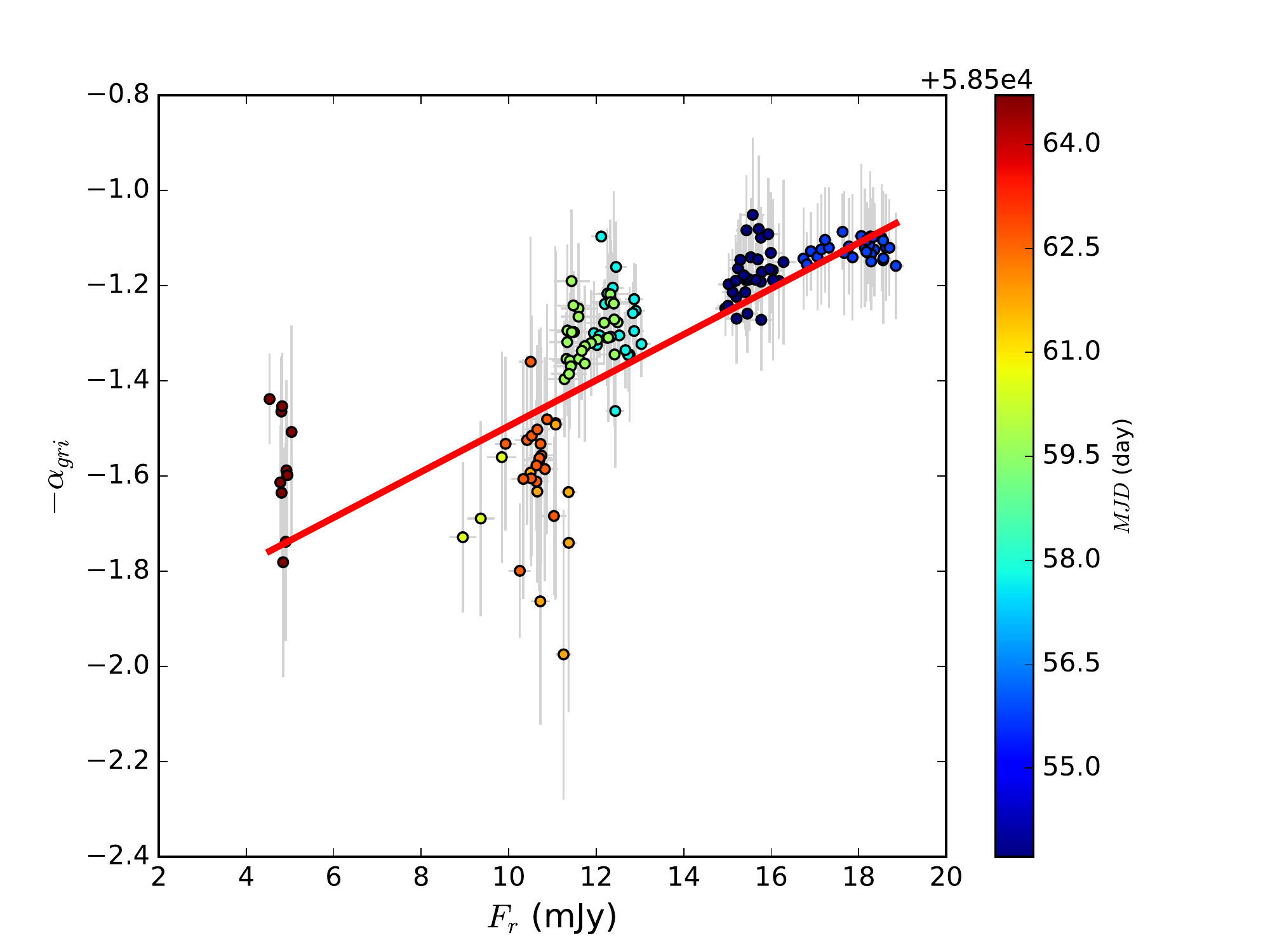}
\includegraphics[angle=0,scale=0.29]{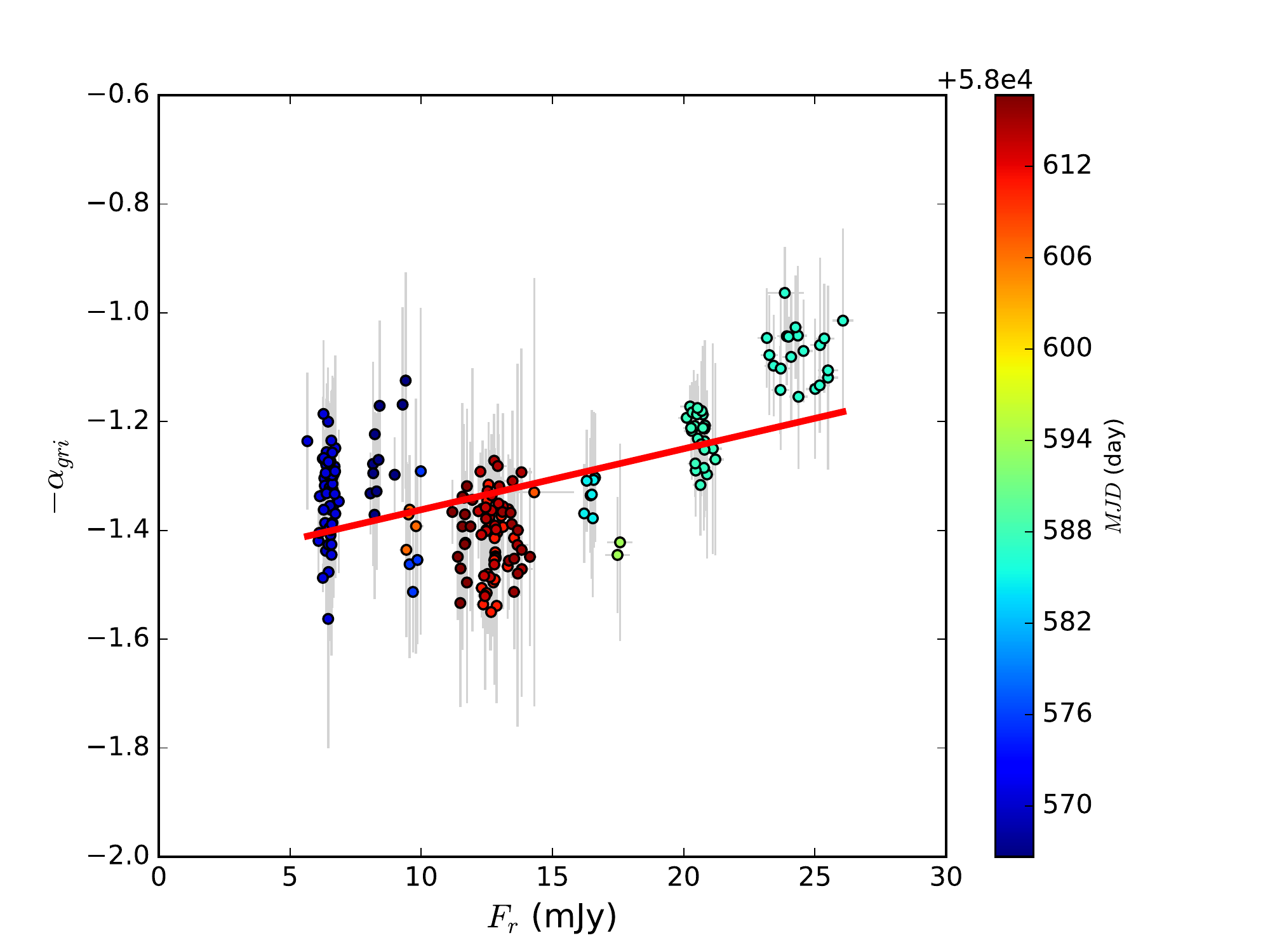}
\includegraphics[angle=0,scale=0.29]{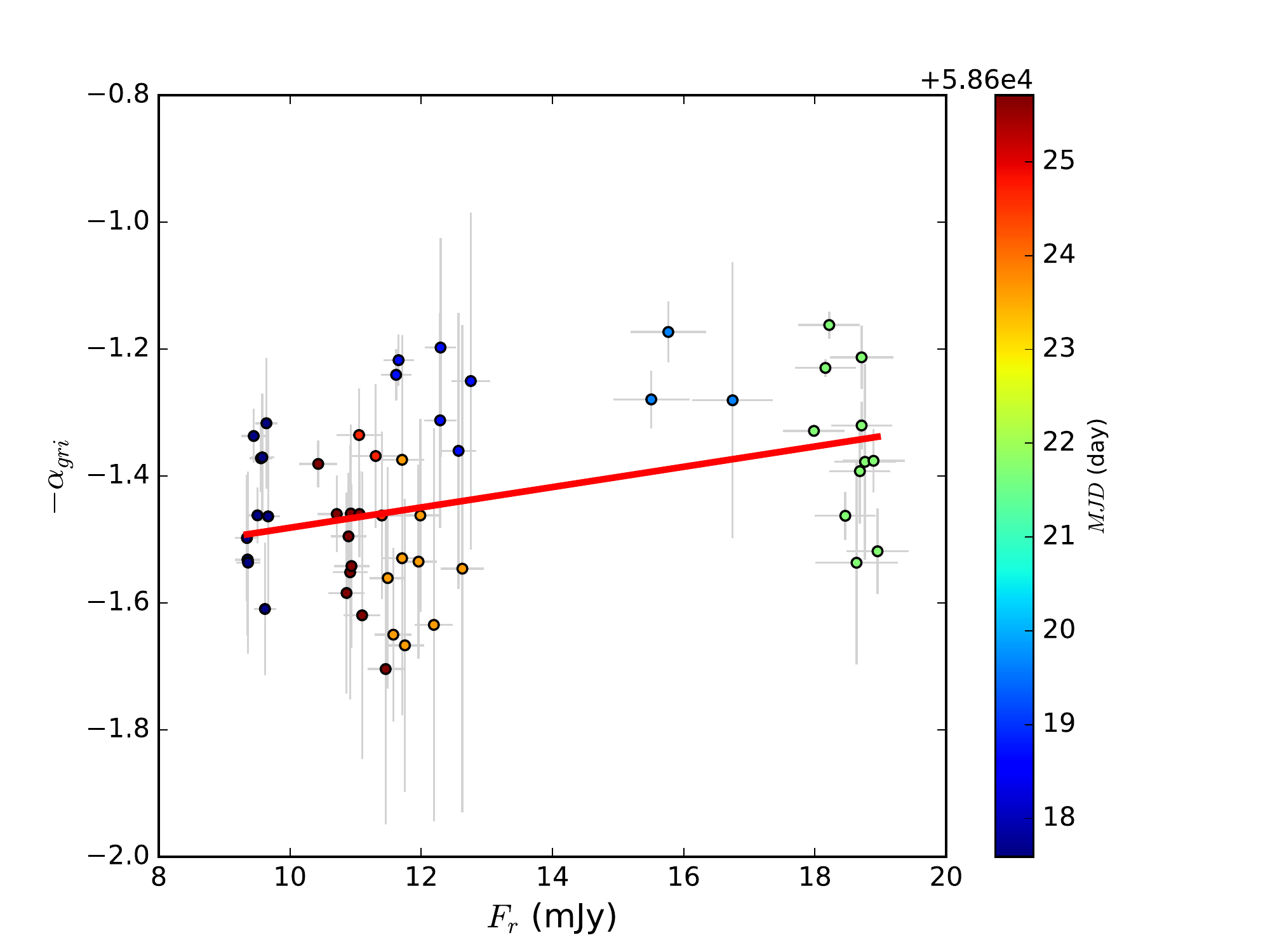}
\includegraphics[angle=0,scale=0.29]{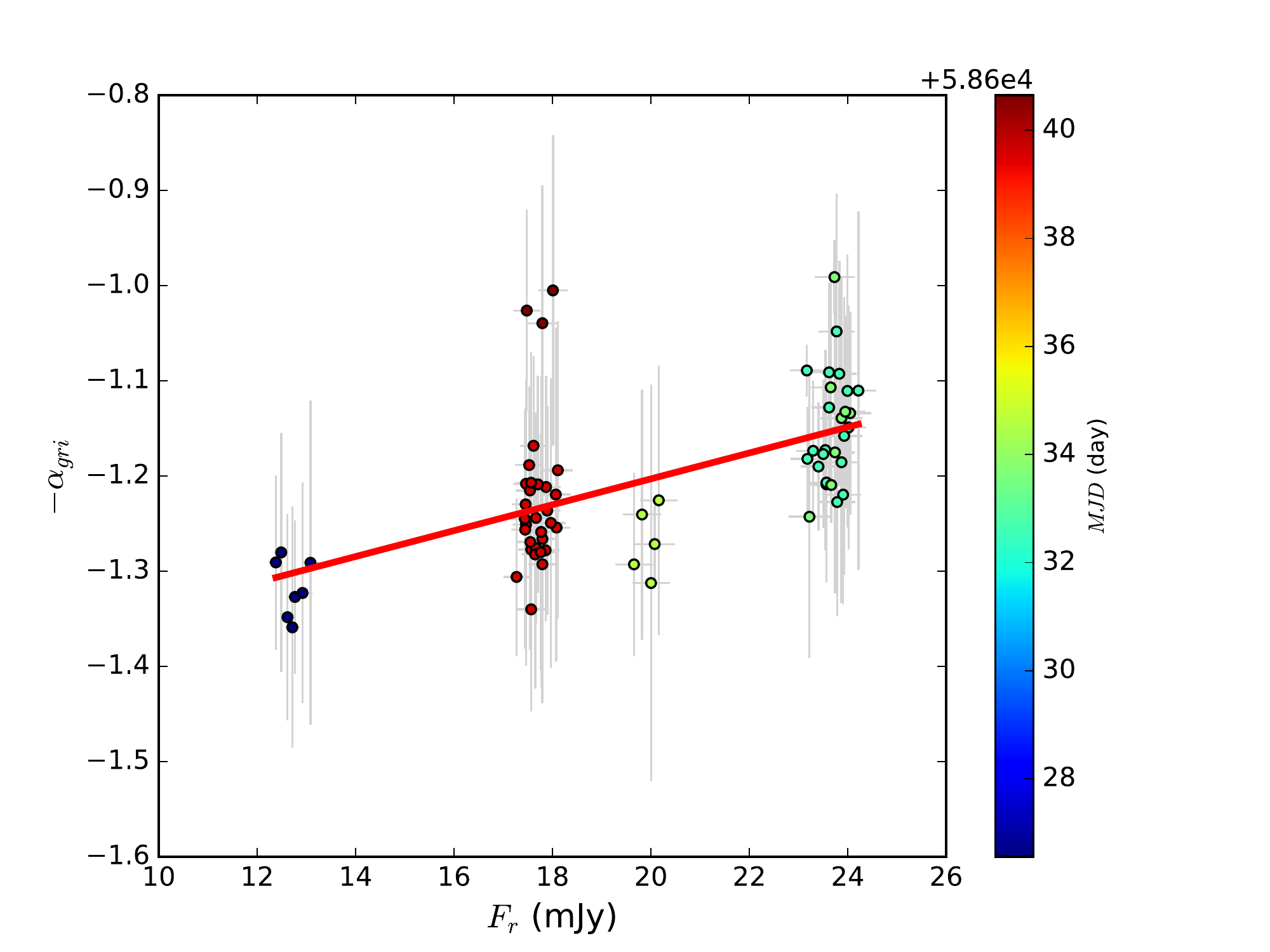}
\caption{The correlations between optical spectral index $\alpha_{\rm gri}$ and $r$ band flux $F_{\rm r}$ for prominent sub-flares. The red solid lines are the results of WLS fitting. The time of 12 panels in turn corresponds to the time of 12 sub-flares in Fig. 8. \label{fig7}}
\end{center}
\end{figure}

\begin{figure}[htb!]
\begin{center}
\includegraphics[angle=0,scale=0.55]{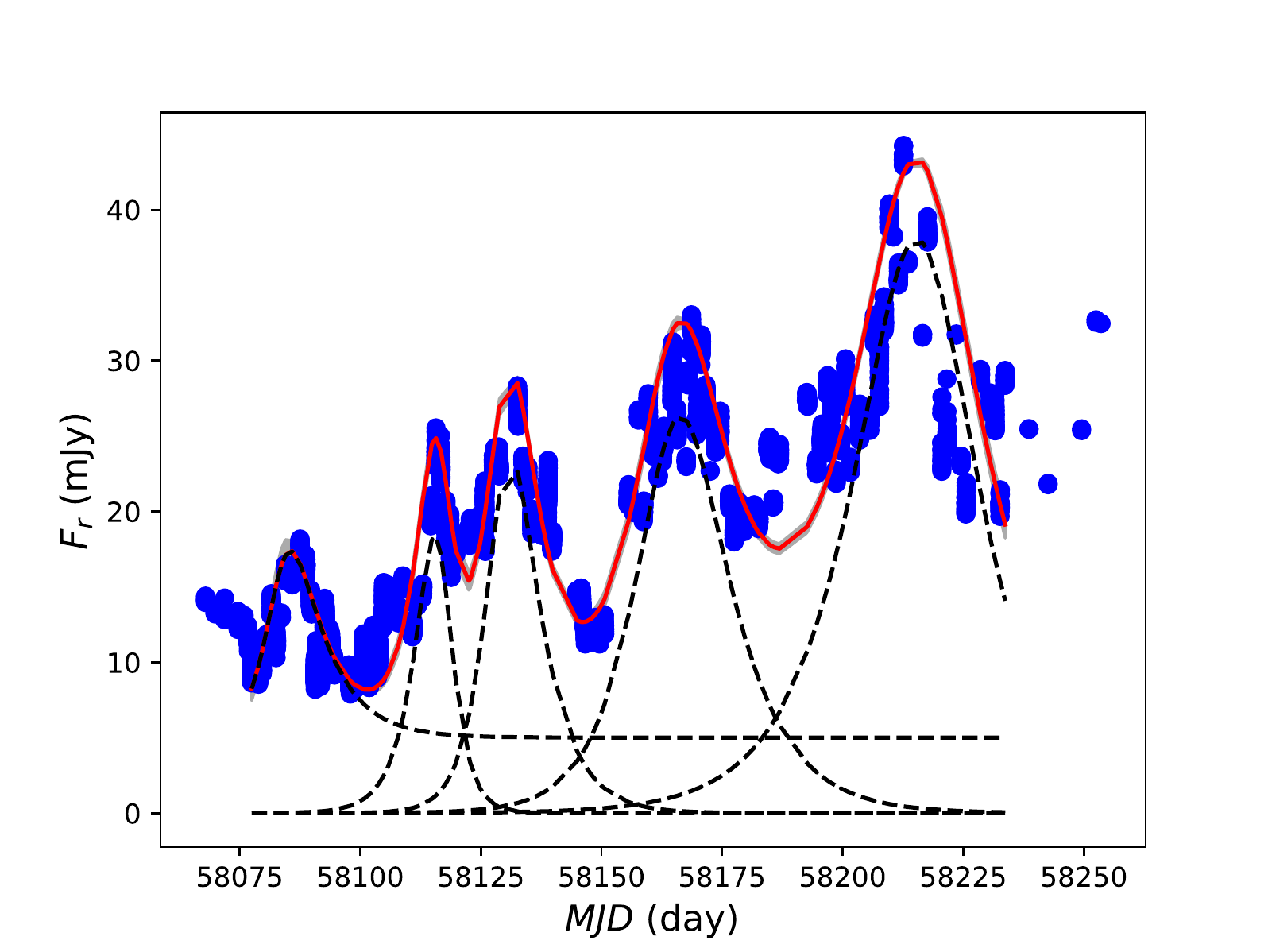}
\includegraphics[angle=0,scale=0.55]{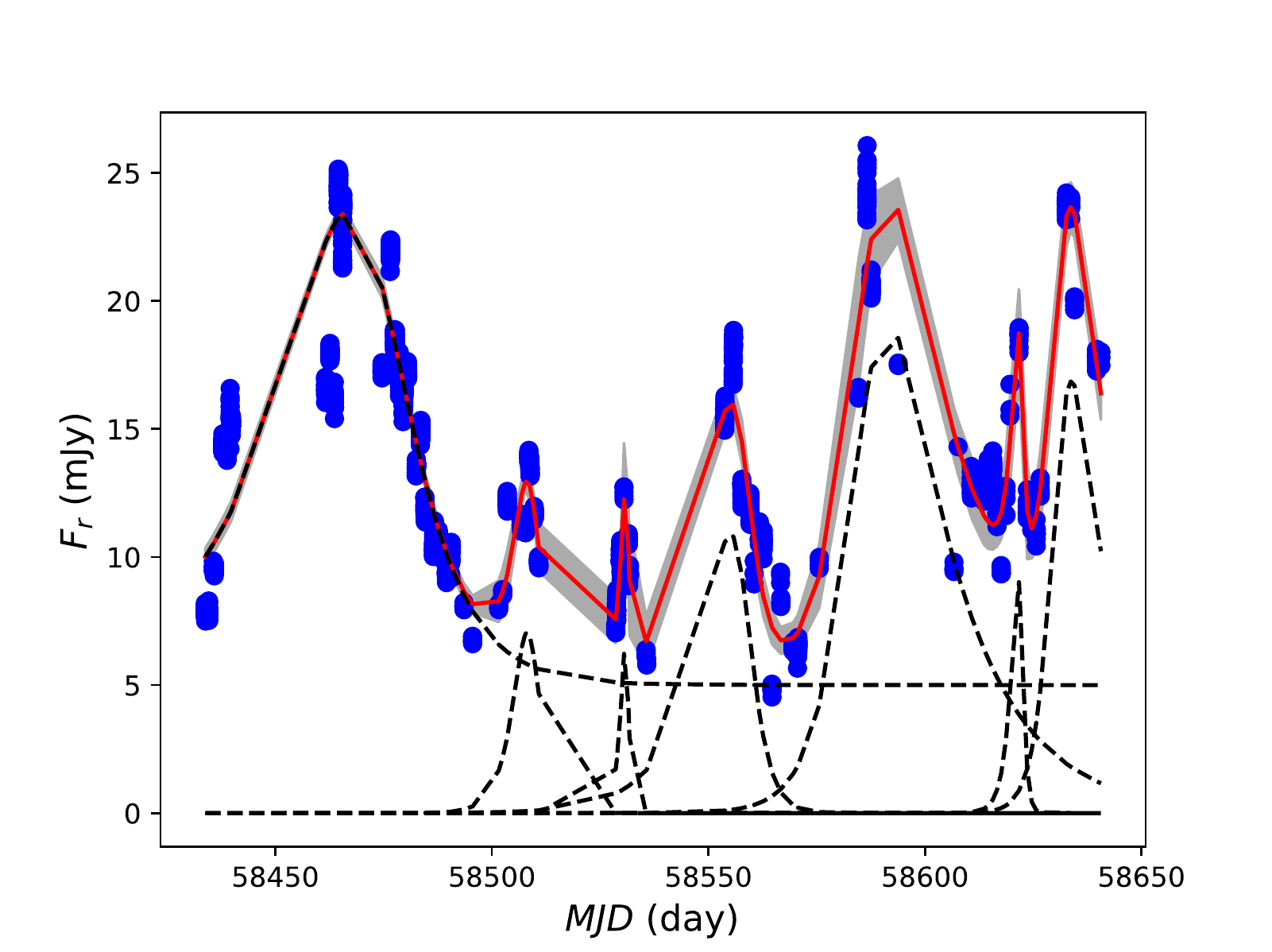}
\caption{The fitting of light curves. The blue circles stand for observed data points. The red solid curves are the background flux ($F_{\rm c}=5$ mJy) plus the flare flux. The black dashed lines represent the background flux and the fitting single flare, respectively. The shadow areas are confidence interval of 3$\sigma$. \label{fig8}}
\end{center}
\end{figure}

To further explore the origins of the relationships between optical index and flux, all prominent sub-flares are analysed in detail. The correlations between $\alpha_{\rm gri}$ and $F_{\rm r}$ are seen in Fig. 7. The results of WLS fitting reveal that there are strong/moderate correlations between $\alpha_{\rm gri}$ and $F_{\rm r}$, except for a weak correlation of the flare from 58617 to 58626. Most of flares have strong FWB trends with different slopes, whereas it should be noted that the slopes both from 58119 to 58150 and from 58180 to 58235 are very small, i.e., a weak FWB trend (Table 3 and Fig. 7). The following function is used to fit the light curves of sub-flares (Abdo et al. 2010b):

\begin{equation}
F(t)=F_{\rm c}+\sum_i F_{\rm 0}\left(e^{\frac{t_{\rm 0}-t}{t_{\rm r}}}+e^{\frac{t-t_{\rm 0}}{t_{\rm d}}}\right)^{-1},~~~~i=1,2,3,...,N,
\end{equation}
where $F_{\rm c}$ represents the background flux, $\frac{F_{\rm 0}}{2}$ is the amplitude of the flares (i.e., $\frac{F_{\rm 0}}{2}=F_{\rm peak}$), $t_{\rm 0}$ is the position of peak, $t_{\rm r}$ and $t_{\rm d}$ are the rise time and the decay time respectively, $N$ is the number of flares. With the help of the non-linear optimization python package (lmfit\footnote{https://lmfit.github.io/lmfit-py/}), the light curves are fitted using the Equation (2). During the modelling process, we only fit these remarkable flares. Although additional small flares added into Equation (2) can get a better fit to light curve, more fitting parameters increase more uncertainties. In order to fit these peaks of flares, most of the peak fluxes are fixed. The lowest flux level is set up as the background flux during the two observed seasons. The fitting results are shown in the Fig. 8 and Table 3. The following equations can be obtained by assuming that $t_{\rm d}\gtrsim t_{\rm cool,obs}$ (cooling timescale), and that $t_{\rm r}$ is related with the observed light crossing timescale (Danforth et al. 2013). The emitting region radius is written as

\begin{equation}
R \lesssim \frac{\mathcal{\delta}ct_{\rm r}}{(1+z)} \simeq 8.4\times 10^{-4}\;{\rm pc}\times\left(\frac{\mathcal{\delta}}{1+z}\right)\left(\frac{t_{\rm r}}{{\rm day}}\right)\,,
 \end{equation}
where the Doppler factor $\delta=10.9$ (Hovatta et al 2009), $c$ is light speed and $z=0.31$. Given that $t_{\rm d}\gtrsim t_{\rm cool,obs}$, we have (Danforth et al. 2013; Ghisellini 2012)
 \begin{equation}
 t_{\rm d} \gtrsim t_{\rm cool,obs} = \frac{3m_{\rm e}c(1+z)}{4\sigma_{\rm T}\mathcal{\delta}u_0'\gamma_{\rm e}}\,,	
 \end{equation}
in which $\gamma_{\rm e}$ is the characteristic random Lorentz factor of electrons, $u_0'=u_B'+u_{\rm rad}'=(1+q)B'^2/(8\pi)$ ($u_B'$ and $u_{\rm rad}'$ are the co-moving energy densities of the magnetic field and the diffuse radiation, respectively) with the Compton dominance parameter $q=u_{\rm rad}'/u_{\rm B}'\simeq L_{\rm IC}/L_{\rm syn}$, $\sigma_{\rm T}$ is the Thomson scattering cross section and $m_{\rm e}$ is electronic mass. The observed frequency of the synchrotron radiation is expressed as (Equation 8.25 from Ghisellini 2012; Yan et al. 2018)
 \begin{equation}
 \nu_{\rm syn,obs} =\frac{4}{3}\nu_{\rm L}\frac{\gamma_e^2\mathcal{\delta}}{(1+z)}=\frac{4}{3}\frac{eB'\gamma_e^2\mathcal{\delta}}{2\pi m_{\rm e}c(1+z)} =\frac{0.212\;eB'\gamma_e^2\mathcal{\delta}}{(1+z)m_{\rm e}c},
 \end{equation}
where $e$ is electronic charge and $B'$ is the magnetic field strength in the emitting region. If the optical radiation is produced by the synchrotron radiation, then the magnetic field strength can be obtained by substituting $\gamma_e$ of Equation (5) into Equation (4):
\begin{eqnarray}
B' &\gtrsim& (0.212m_{\rm e}ce)^{1/3}\times\left(\frac{6\pi}{\sigma_{\rm T}}\right)^{2/3}\left[\frac{(1+z)}{\mathcal{\delta}\nu_{\rm syn,obs}}\right]^{1/3}[(1+q)t_{\rm d}]^{-2/3}
\nonumber\\
&=& 1.29\times10^8\;{\rm G}\times \delta^{-1/3}(1+q)^{-2/3}\times\left(\frac{\nu_{\rm syn,obs}}{{\rm Hz}}\right)^{-1/3}\left(\frac{t_{\rm d}}{{\rm s}}\right)^{-2/3}\left(1+z\right)^{1/3}\,.                                                                                                           
\end{eqnarray}
From Equation (6) and Equation (5), the random Lorentz factor of electrons is 
\begin{eqnarray}
\gamma_{\rm e} &\lesssim& \left(\frac{\sigma_{\rm T}m_{\rm e}c}{6\pi}\right)^{1/3}(0.212e)^{-2/3}\times\left[\frac{(1+z)(1+q)t_{\rm d}}{\mathcal{\delta}}\right]^{1/3}\nu_{\rm syn,obs}^{2/3}
\nonumber\\
&=& 21.07\times10^{-26/3}\times \delta^{-1/3}(1+q)^{1/3}\times\left(\frac{\nu_{\rm syn,obs}}{{\rm Hz}}\right)^{2/3}\left(\frac{t_{\rm d}}{{\rm s}}\right)^{1/3}\left(1+z\right)^{1/3}\,.
\end{eqnarray} 
The apparent flare luminosity is that $L= 4\pi d_{\rm L}^2F_{\rm peak} \;{\rm erg\,s^{-1}}$ ($d_{\rm L}$ is the luminosity distance), and the co-moving energy density of the synchrotron radiation (Danforth et al. 2013)
\begin{eqnarray}
u_{\rm syn}' &=& \frac{L}{4\pi c\mathcal{\delta}^4R^2} \gtrsim \frac{1}{4\pi c^3}\times\frac{(1+z)^2L}{\mathcal{\delta}^6t_{\rm r}^2}
\nonumber\\
&=& 2.95\times10^{-33}\;{\rm erg\,cm^{-3}}\times \delta^{-6}\left(\frac{t_{\rm r}}{{\rm s}}\right)^{-2}\times\left(\frac{L}{{\rm erg\,s^{-1}}}\right)\left(1+z\right)^{2}\,.                                                             
\end{eqnarray}
From Danforth et al. (2013), the number of electrons contributing to the optical emission is
\begin{eqnarray}
N_{\rm e}&\simeq& 6\pi L/(\sigma_{\rm T}c\mathcal{\delta}^4B'^2\gamma_{\rm e}^2)\\ \nonumber
          &\simeq& 2.74\times 10^{13}\times \delta^{-8/3}(1+q)^{2/3}\times\left(1+z\right)^{-4/3}\left(\frac{L}{{\rm erg\,s^{-1}}}\right)\left(\frac{\nu_{\rm syn,obs}}{\rm Hz}\right)^{-2/3}\left(\frac{t_{\rm d}}{{\rm s}}\right)^{2/3},
\end{eqnarray} 
the co-moving energy density of the electrons
\begin{eqnarray}
 u_{\rm e}'&\simeq& 3N_{\rm e}\gamma_{\rm e}m_{\rm e}c^2/(4\pi R^3)\\ \nonumber                                                         
           &\simeq& 1.956\times 10^{-7}\;{\rm erg\,cm^{-3}}\times \left(\frac{R}{{\rm cm}}\right)^{-3}\times N_{\rm e} \gamma_{\rm e}\,, 
\end{eqnarray}
and the ratio of co-moving electron to synchrotron radiation energy density
\begin{equation}
	\frac{u_{\rm e}'}{{u_{\rm syn}'}}\propto\frac{t_{\rm d}}{{t_{\rm r}}}(1+q)\,.
\end{equation}
For the target source, the Compton dominance parameter $q$ is taken to be 1, and the $r$ band frequency is $c/6231{\rm\AA}$. The flux variability is measured by the fractional variability parameter $F_{\rm var}$ (Vaughan et al. 2003; Aleksic et al. 2015):
 \begin{equation}
 F_{\mathrm{var}} = \sqrt{\frac{S^2 - <\sigma_{\mathrm{err}}^2 >}{<F>^2}}\,,
 \end{equation}
 where $S$ is the standard deviation of $F$, $<\sigma_{\mathrm{err}}^2 >$ is the mean value of squared error and $<F>$ is the average flux. The uncertainty of $F_{\rm var}$ (Poutanen et al. 2008; Vaughan et al. 2003; Aleksic et al. 2015) is calculated by 
 \begin{equation}
 \Delta F_{\mathrm{var}} = \sqrt{F^{2}_{\mathrm{var}} +
    err(\sigma^{2}_{\mathrm{NXS}})} -F_{\mathrm{var}}\,, \nonumber 
  \end{equation}
  \begin{equation}
 err(\sigma^{2}_{\mathrm{NXS}}) = \sqrt{\left(\sqrt{\frac{2}{N}}\frac{<\sigma^{2}_\mathrm{err}>}{<F>^{2}}\right)^{2}  +
    \left(\sqrt{\frac{<\sigma^{2}_\mathrm{err}>}{N}}\frac{2F_{\mathrm{var}}}{<F>}\right)^{2}} ~. \nonumber
  \end{equation}

The fitting values of $t_{\rm r}$, $t_{\rm d}$ and $F_{\rm 0}$ constrain the $R$, $B'$, $\gamma_{\rm e}$, $u_{\rm syn}'$, $N_{\rm e}$, $u_{\rm e}'$ and $u_{\rm e}'/u_{\rm syn}'$ by using the above equations (Table 3). The upper limits of emitting region range from 0.009 pc to 0.135 pc with an average value of 0.043$\pm$0.01 pc, the lower limits of magnetic field strength from 0.04 G to 0.43 G with an average value of 0.13$\pm$0.03 G, the upper limits of random Lorentz factor from 6001 to 19254 with an average value of 13272$\pm$1153, the lower limits of co-moving energy density of the synchrotron radiation from $3\times10^{-5}$ erg~${\rm cm}^{-3}$ to $2.1\times10^{-3}$ erg~${\rm cm}^{-3}$ with an average value of $(7.2\pm1.8) \times 10^{-4}$ erg~${\rm cm}^{-3}$, the co-moving energy densities of the electrons from $\sim9.4\times10^{-5}$ erg~${\rm cm}^{-3}$ to $\sim2.5\times10^{-2}$ erg~${\rm cm}^{-3}$ with an average value of $\sim(5.5\pm1.9) \times 10^{-3}$ erg~${\rm cm}^{-3}$, and the ratios of co-moving electron to synchrotron radiation energy density from $\sim1.98$ to $\sim20.14$ with an average value of $\sim7.71\pm1.7$. The number of electrons ranges from $\sim9.5\times10^{49}$ to $\sim4.9\times10^{51}$ with an average value of $\sim(1.66\pm0.4)\times10^{51}$, the fractional variability parameter $F_{\rm var}$ from 0.11 to 0.45 with an average value of $0.24\pm$0.02, the average flux of each sub-flare from 8.8 mJy to 26.4 mJy with an average value across all sub-flares of 15.73$\pm$1.5 mJy, and the average spectral index of each sub-flare from 1.17 to 1.51 with an average value across all sub-flares of 1.3$\pm$0.03.  

\begin{figure}[htb!]
\begin{center}
\includegraphics[width=15cm,height=12cm]{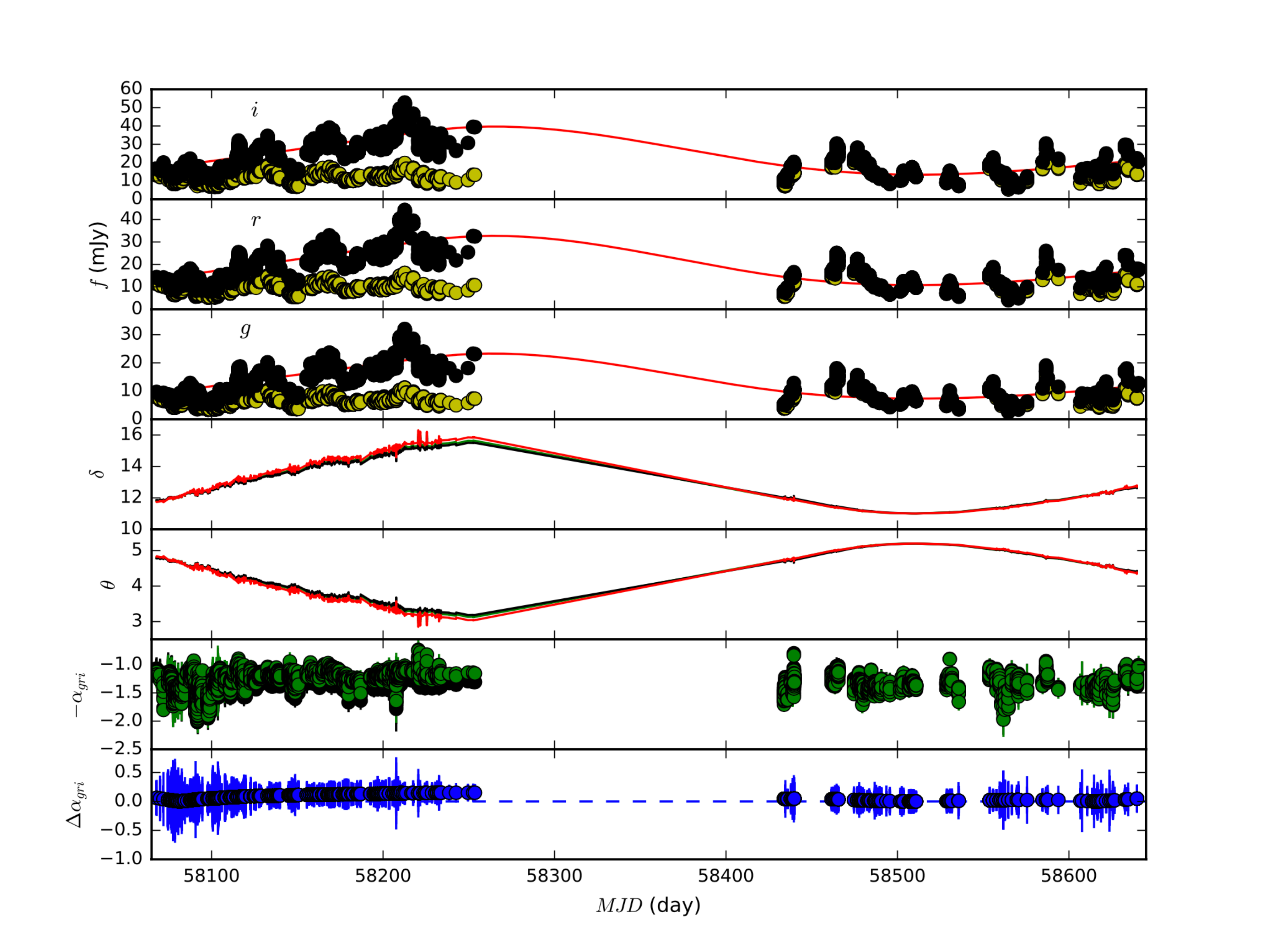}
\caption{The variations from light curves and spectral index for 60-days binning data. The fourth and fifth panels represent the Doppler factor and the angle of the line of sight relative to the jet in different bands. The meanings of the other panels are same as that of Fig. 5.\label{fig9}}
\end{center}
\end{figure}

The correlations between the slope ($k$) and one of the parameters ($R$, $B'$, $\gamma_{\rm e}$, $u_{\rm syn}'$, $N_{\rm e}$, $u_{\rm e}'$ and $u_{\rm e}'/u_{\rm syn}'$) are analysed. The results of Pearson correlation\footnote{https://docs.scipy.org/doc/scipy/reference/generated/scipy.stats.pearsonr.html} reveal that there are no significant correlations between them (the probability of chance correlation $0.2<p<0.9$), except for the relation between $k$ and $N_{\rm e}$ (the correlation coefficient $r=-0.56$, $p=0.057$). After the influence from flare luminosity is considered, the partial correlation analysis shows that the correlation between $k$ and $N_{\rm e}$ disappears ($r=0.01$, $p=0.95$), since $k$ depends on the flare luminosity ($r=-0.6$, $p=0.04$). We also use a Spearman rank-order correlation\footnote{https://docs.scipy.org/doc/scipy/reference/generated/scipy.stats.spearmanr.html\#scipy.stats.spearmanr} to analyse these correlations because the Spearman correlation is a nonparametric measure of the monotonicity of the relationship between two datasets. The results indicate that the relationships of both $k$ versus $B'$ and $k$ versus $\gamma_{\rm e}$ could have tentative significant correlations (Spearman correlation coefficient $r=0.56$, significant level $p=0.06$ and $r=-0.56$, $p=0.06$). If the significant level is less than 0.05, the correlation is significant at a 95\% confidence level. The relationships between the fractional variability parameter ($F_{\rm var}$) and one of the parameters ($<F>$, $B'$, $\gamma_{\rm e}$, $u_{\rm syn}'$, $N_{\rm e}$, $u_{\rm e}'$ and $u_{\rm e}'/u_{\rm syn}'$) have no significant correlations, except for a tentative significant correlation between $F_{\rm var}$ and $R$ (Spearman correlation coefficient $r=0.5$ and significant level $p=0.09$).

In the last part of the section, we continue to take longer binning data to track flux variability in the long-term trend by using the method of Villata et al. (2004), as shown above. The values of $\Delta \alpha_{\rm gri}$ within the error range approach zero when the 60-days binning is selected (see Fig. 9). Thus, the long-term trend of 60-day binning data could be tracked by the variations of Doppler factor because the variations of Doppler factor could not make the spectral index change. The minimum values from spline interpolations are 13.4 mJy, 10.8 mJy and 7.4 mJy respectively for $i$, $r$ and $g$ bands. We assume the minimum values as constant base-level fluxes similar to Raiteri et al. (2017), which correspond to the Lorentz factor $\Gamma=10.3$ and the orientation $\theta=5.2^{\circ}$ (Hovatta et al. 2009). According to $F_{\rm const}=F_{\nu}\left(\frac{\delta_{\rm const}}{\delta_{\nu}}\right)^{2+\alpha}$ (Raiteri et al. 2017), the Doppler factor can be written as
  \begin{equation}
 \delta_{\rm \nu}=\delta_{\rm const}\left(\frac{F_{\rm \nu}}{F_{\rm const}}\right)^{\frac{1}{2+\alpha}}=\delta_{\rm const}C_j(t)^{\frac{1}{2+\alpha}}\,.
  \end{equation}
 Assuming that the Doppler factor variations are caused by the orientation variations, the orientation variations can be estimated by taking into account that the orientation $\theta={\rm arccos}[(1-(\delta \Gamma)^{-1})\beta^{-1}]$ and the bulk velocity $\beta=\sqrt{1-\Gamma^{-2}}$ (Raiteri et al. 2017; Urry \& Padovani 1995). The variations of Doppler factor and orientation can track the long-term trend of 60-days binning data well (Fig. 9). The Doppler factor changes between 11 and 16, and the orientation between $5.2^\circ$ and $2.95^\circ$. 
 
 We should notice that the Doppler factor keeps constant ($\delta=10.9$) when estimating the parameters in Table 3. After the long-term trend of 60-days binning data is removed, the sub-flares in the light curve are fitted using the Equation (2). The model parameters are estimated using above Equations (3)-(14) in which the Doppler factor $\delta=10.9$ (Table 4). The upper limits of emitting region range from 0.009 pc to 0.128 pc with an average value of 0.039$\pm$0.01 pc, the lower limits of magnetic field strength from 0.048 G to 0.42 G with an average value of 0.13$\pm$0.03 G, the upper limits of random Lorentz factor from 6043 to 17812 with an average value of 12928$\pm$1053, the lower limits of co-moving energy density of the synchrotron radiation from $2.8\times10^{-5}$ erg~${\rm cm}^{-3}$ to $2.6\times10^{-3}$ erg~${\rm cm}^{-3}$ with an average value of $(7.2\pm2.2) \times 10^{-4}$ erg~${\rm cm}^{-3}$, the co-moving energy densities of the electrons from $\sim7.4\times10^{-5}$ erg~${\rm cm}^{-3}$ to $\sim1.6\times10^{-1}$ erg~${\rm cm}^{-3}$ with an average value of $\sim(1.6\pm1.2) \times 10^{-2}$ erg~${\rm cm}^{-3}$, and the ratios of co-moving electron to synchrotron radiation energy density from $\sim1.89$ to $\sim59$ with an average value of $\sim10.4\pm4.3$. The number of electrons ranges from $\sim1\times10^{50}$ to $\sim1.9\times10^{51}$ with an average value of $\sim(9.3\pm1.6)\times10^{50}$, the fractional variability parameter $F_{\rm var}$ from 0.11 to 0.45 with an average value of $0.24\pm$0.02, the average flux of each sub-flare from 7.5 mJy to 13.7 mJy with an average value across all sub-flares of 10.34$\pm$0.5 mJy, and the average spectral index of each sub-flare from 1.24 to 1.54 with an average value across all sub-flares of 1.33$\pm$0.02. After the long-term trend of 60-days binning data is removed, there are no significant correlations between $k$ (or $F_{\rm var}$) and one of the parameters ($R$, $B'$, $\gamma_{\rm e}$, $u_{\rm syn}'$, $u_{\rm e}'$ and $u_{\rm e}'/u_{\rm syn}'$) for either Pearson correlation or Spearman correlation. The significant correlation between $k$ and $N_{\rm e}$ is also due to the influence from flare luminosity.
  
\section{Discussion}

In the beginning of this section, we compare our results with previous works, and then discuss the origins concerning color behaviours and flux variability in the long-term timescales. The long-term $r$ band light curves show that the target source brightens from $14^{\rm m}.16$ to $12^{\rm m}.29$ together with five prominent sub-flares in the first observed season, and first becomes fainter from $12^{\rm m}.89$ to $14^{\rm m}.76$ and then brightens from $14^{\rm m}.76$ to $12^{\rm m}.94$ with seven prominent sub-flares in the second observed season. The average values are $13.29^{\rm m}\pm0.01$ and $13.60^{\rm m}\pm0.01$ respectively for the two observed seasons. The Tuorla Blazar Monitoring Program\footnote{http://users.utu.fi/kani/1m/index.html} (TBMP; Takalo et al. 2008) has observed the target source in the optical $R$ band. Data reduction procedure and interpretation are available at Nilsson et al. (2018). The results of comparison show that the changes of our light curves are consistent with that of the TBMP during the period. After considering the different comparison stars, bands and time, our observation data are in agreement with the archive data from Steward observatory monitoring project\footnote{http://james.as.arizona.edu/\~{}psmith/Fermi/} (Smith et al. 2009). 

We find that for this source the maximum and minimum magnitudes are $15^{\rm m}.5$ and $11^{\rm m}.6$ respectively with an average value of $13^{\rm m}.3$ via checking the light curve of TBMP observed from 2004 to 2019. Simply, we classify the brightness below $13^{\rm m}.46$ (corresponding to 15 mJy) as a low flux state, the brightness above $12^{\rm m}.5$ as a high flux state and the brightness from $13^{\rm m}.46$ to $12^{\rm m}.5$ as a middle flux state by checking the light curve from TBMP. So the target source transitions from a low state to a high state in the first observed season, and from a middle flux state to a low state and then from a low flux state to a middle state in the second observed season. 

For our observations, the time intervals between two adjacent bands range from 22 seconds to 130 seconds with an average value of 46 seconds. The total number of observations is 201 nights including 26973 data points. The average time spans and time resolutions are 3.4 hours and 2.9 minutes per night, respectively. Dai et al. (2013) conducted multicolor optical monitoring for the object from 2004 to 2011. Their observations presented 8661 data points. The multi-band optical observations across eight years provided 5186 data points (Dai et al. 2015). Yuan et al. (2017) reported 4969 sets of photometrical optical observations in the monitoring time from 2000 to 2014. Hu et al. (2014) monitored the source from 2009 to 2014, and obtained 6176 photometric observations. 4854 data points observed from 1994 to 2002 had been collected in the $UBVRI$ bands (Raiteri et al. 2003). For the target source, the above large samples span longer-term timescales but get less data points compared to our sample with high dense sampling of our observations. In addition, the optical $g$, $r$ and $i$ bands are exposed in turn, and the time intervals between two adjacent bands less than 2.2 minutes, which indicates that our observations are quasi-simultaneous multicolor optical monitoring. Consequently, our observations gain a very large quasi-simultaneous multicolor optical sample for the target source.

Our results that a strong FWB trend at a low flux state and then a weak FWB trend at a higher flux state are reported in the first time for the target source. For the source, the previous works (Poon et al. 2009; Wu et al. 2007; Dai et al. 2015; Dai et al. 2013; Raiteri et al. 2003; Ghisellini et al. 1997; Villata et al. 2000, 2008; Chandra et al. 2011; Hu et al. 2014; Stalin et al. 2006; Agarwal et al. 2016; Bhatta et al. 2016; Hong et al. 2017; Zhang et al. 2018) did not obviously see the different color trends in low and high flux states, which caused that the color behaviours from the previous works did not consider the different trends in low and high flux states and were only analysed by using a single trend. If the previous works had enough sampling rate and also defined the break flux, such trends or different trends could be found. Then the above results were not observed in the previous works in part because either the spectral trends were peculiar or there was not enough sampling rate. The results that a weak FWB trend at a low flux state and then a strong FWB trend at a higher flux state are also found.

The brightness contributions from host galaxy can be neglected for the source due to the very low brightness of its host galaxy compared to the source, i.e., the optical spectral behaviours can not be explained by the host galaxy contributions, even at a low flux state. Generally, a change of the viewing angle on a ``convex" spectrum could be the mechanism to account for a mild BWB trend or an achromatic color change in the long-term timescales (Villata et al. 2002, 2004; Papadakis et al. 2007; Larionov et al. 2010). Assuming that the achromatic color trends or mild FWB (BWB) trends in the long-term timescales are due to a change of the viewing angle on a ``convex'' spectrum, then a weak FWB trend should be significantly improved to a middle/strong FWB trend after the long-term trends are removed. The divergence between our results and such an assumption likely indicates that the optical spectrum could not be a ``convex'' spectrum but a power law spectrum. Our results display that the relationships between $r$ band flux and $g$ (or $i$) band flux can be linearly fitted well, and that after the long-term trends are removed, the FWB trends between $\alpha_{\rm gri}$ and $F_{\rm r}$ are no significant improvement. Additionally, the frequencies of optical $g$, $r$ and $i$-bands are neighbouring ($g$ band: $6.2\times10^{14}$ Hz; $r$ band: $4.8\times10^{14}$ Hz; $i$ band: $3.9\times10^{14}$ Hz). So the variations of Doppler factor are unlikely to make the spectral index (or color) change for a power law spectrum, i.e., the variations of Doppler factor could not be as the main mechanism to explain the FWB trends for the target source during the period of our observations. The different blazars and observed periods may cause that our results are different from above previous works. Meanwhile, for S5 0716+714, it is worth to analyse the effect of Doppler factor on the optical spectral behaviours in longer-term timescales.

For some FSRQs, the RWB trend was found at a low flux state due to a blend of accretion disk and jet components (Gu et al. 2006). The saturation effect of constant color appears with brightening when the jet-disk contributions are equal, and the BWB trend at a high flux state when the jet emission is dominant (Isler et al. 2017). If the thermal component from accretion disk dominates the total radiation, accretion disk model could also produce the BWB trend (Liu et al. 2016; Gu \& Li 2013; Li \& Cao 2008). For the BL Lac object S5 0716+714, given its high power and lack of signs for ongoing accretion or surrounding gas, it is very hard that the thermal component from accretion disk dominates the total radiation. Then a strong FWB trend or a weak FWB trend at a low flux state could not be explained by accretion disk components. The SWB trends at a low flux state are also not found from our results. Therefore, the accretion disk components are impossible to interpret our optical spectral behaviours. Alternatively, the jet dominated mechanisms are presented to explain the optical spectral behaviours.

For jet dominant emission, the changes of color trends or spectral index trends are likely dominated by the particle acceleration and cooling mechanisms; the BWB trends or FWB trends were thought to be due to electrons being accelerated to preferentially higher energies before radiative cooling, i.e., the spectral evolution of the accelerated electrons to higher intensities and higher spectral frequency; the redder when fainter (RWF) trends or steeper when fainter (SWF) trends occur when the highest-energy electrons suffer a stronger radiative cooling or escape cooling that causes the spectral evolution of the electrons to lower intensities and lower spectral frequency (Isler et al. 2017). If the electrons in the emission zone keep the spectral shape unchanged and the source brightens/darkens by changing the number of electrons, then the color trends or spectral index trends keep unchanged when the target source becomes bright or dark, e.g., the acceleration and radiative cooling keep dynamic balance to the electrons. 

For our results in the first observed season, a strong FWB trend at a low flux state ($F_{\rm r}<15$ mJy) is due to the spectral evolution of most of the accelerated electrons to higher intensities and higher spectral frequency. The spectral evolution of partial accelerated electrons to higher intensities and higher spectral frequency while the other electrons keep the spectral shape unchanged or other spectral shapes. The different spectral evolutions from the electrons cause a weak FWB trend at a higher flux state ($F_{\rm r}>15$ mJy). The reverse explanation is proposed to interpret the results of a weak FWB trend at a low flux state and then a strong FWB trend at a higher flux state in the second observed season. The discrepancy of results from the two observed seasons could be owing to different transitions in states that the target source transitions from a low state to a high state in the first observed season, and from a middle flux state to a low state and then from a low flux state to a middle state in the second observed season.

In oder to further explore the spectral behaviours and flux variability, all prominent sub-flares are analysed in detail. Most of electrons have similar strong FWB-slopes leading to a strong FWB trend at a low flux state in the first observed season (Fig. 7, 8 and Table 3; $r=0.7$, $P<10^{-4}$ and $k=0.077$ from 58067 to 58079; $k=0.067$ from 58079 to 58100 corresponding to the first flare). For a weak FWB trend at a higher flux state in the first observed season, the second and the fourth flares have similar strong FWB-slopes (Fig. 7, 8 and Table 3; $k=0.019$ and $k=0.012$) while the third and the fifth flares have a weak FWB-slope (Fig. 7, 8 and Table 3; $k=0.0055$ and $k=0.0061$). After the superposition of the different slopes, a weak FWB trend appears. In the second observed season, the seven sub-flares show the different FWB-slopes (Fig. 7, 8 and Table 3). For flux below 15 mJy, the superposition of the different FWB-slopes can be as the reason of a weak FWB trend at a low flux state, and for flux above 15 mJy, the superposition of similar strong FWB-slopes as the reason of a strong FWB trend at a higher flux state. The particle acceleration and cooling mechanisms can also explain the spectral behaviours for these sub-flares in the short-term timescales as discussed above. Notice that the seventh and the eighth sub-flares with a lower flux state have strong FWB trends, which indicates that the disk contributions should be ignored because the SWB (RWB) trends are not discovered in such low flux state. In this case, it is rational that the optical radiation is produced by synchrotron radiation. 

For the sub-flares, we estimate the limits or approximate values of $R$, $B'$, $\gamma_{\rm e}$, $u_{\rm syn}'$, $u_{\rm e}'$, $N_{\rm e}$ and $u_{\rm e}'/u_{\rm syn}'$ through assuming that $t_{\rm d}\gtrsim t_{\rm cool,obs}$, that the optical radiation is produced by synchrotron radiation, and that $t_{\rm r}$ is related with the observed light crossing timescale. The results of Spearman rank-order correlation indicate that the relationships between $k$ and $B'$ and between $F_{\rm var}$ and $R$ could have tentative significant correlations. After the long-term trend of 60-days binning data is removed, the tentative significant correlations both disappear. The relationships between $k$ (or $F_{\rm var}$) and other parameters have no significant correlations. The limited statistical numbers (12 sub-flares) and having no very dense sampling in the second season may cause no significant correlations. Moreover, it should be noted that we estimate the values of limits or approximate values rather than the accurate values. A stochastic or complicated process may also explain the lack of significant correlations.

Finally, a scenario based on the variations of Doppler factor is proposed to interpret the flux variability in the long-term trend. For a power law spectrum, if the variations of Doppler factor cause the variations of long-term trend in the light curve, the $\Delta \alpha_{\rm gri}$ should be to come near zero after the long-term trend is removed because the variations of Doppler factor are unlikely to make the spectral index change. However, our results from 4-days binning data to 30-days binning data do not agree with such an assumption, i.e., the long-term trends from 4-days binning data to 30-days binning data could not be tracked by the variations of Doppler factor well. We continue to take longer binning data to track flux variability in the long-term trend. The values of $\Delta \alpha_{\rm gri}$ within the error range approach zero when the 60-days binning is selected, which indicates that the long-term trend of 60-days binning data could be tracked by the variations of Doppler factor well. Assuming that the Doppler factor variations are caused by the orientation variations, the flux variability in the long-term trend could be explained by the orientation variations. Therefore, our results suggest that the jet has a bent trajectory or helical path. Such a scenario is consistent with the observed results from VLBI or VLBA (Bach et al. 2005; Nesci et al. 2005; Larionov et al. 2013; Rani et al. 2014, 2015). In addition, twelve prominent sub-flares superimposed in the long-term trend may be related with shock/turbulence in jet (e.g., Marscher \& Gear 1985; Marscher et al. 2008, 2014; Narayan \& Piran 2012; Saito et al. 2015; MAGIC Collaboration et al. 2019). 

\section{Summary}
We continuously monitored the blazar in optical $gri$-bands from Nov. 10, 2017 to Jun. 06, 2019. The total number of observations is 201 nights including 26973 data points that represent a very large quasi-simultaneous multicolor optical sample. The average time spans and time resolutions are 3.4 hours and 2.9 minutes per night, respectively. The main conclusions are the following.

(i) The target source brightens from $14^{\rm m}.16$ to $12^{\rm m}.29$ together with five prominent sub-flares in the first observed season (from Nov. 10, 2017 to May 15, 2018), and first becomes fainter from $12^{\rm m}.89$ to $14^{\rm m}.76$ and then brightens from $14^{\rm m}.76$ to $12^{\rm m}.94$ with seven prominent sub-flares in the second observed season (from Nov. 11, 2018 to Jun. 06, 2019). The average values are $13.29^{\rm m}\pm0.01$ and $13.60^{\rm m}\pm0.01$ respectively for the two observed seasons, and the overall flux variability amplitudes both about $1^{\rm m}.87$. 

(ii) A strong FWB trend at a low flux state and then a weak FWB trend at a higher flux state in the first observed season, and a weak FWB trend at a low flux state and then a strong FWB trend at a higher flux state in the second observed season are both reported. Most of sub-flares show the strong FWB trends, except for two flares with a weak FWB trend. These different sub-flares have different FWB-slopes. The variations of Doppler factor and accretion disk components are unlikely to be as the main mechanisms to explain the optical color behaviours for the target source during the period of our observations. The particle acceleration and cooling mechanisms together with the superposition of the different FWB-slopes from sub-flares are likely to explain the optical color behaviours.

(iii) After the long-term trend is removed, for these sub-flares, the upper limits of emitting region range from 0.009 pc to 0.128 pc with an average value of 0.039$\pm$0.01 pc, the lower limits of magnetic field strength from 0.048 G to 0.42 G with an average value of 0.13$\pm$0.03 G, the upper limits of random Lorentz factor from 6043 to 17812 with an average value of 12928$\pm$1053, the lower limits of co-moving energy density of the synchrotron radiation from $2.8\times10^{-5}$ erg~${\rm cm}^{-3}$ to $2.6\times10^{-3}$ erg~${\rm cm}^{-3}$ with an average value of $(7.2\pm2.2) \times 10^{-4}$ erg~${\rm cm}^{-3}$, the co-moving energy densities of the electrons from $\sim7.4\times10^{-5}$ erg~${\rm cm}^{-3}$ to $\sim1.6\times10^{-1}$ erg~${\rm cm}^{-3}$ with an average value of $\sim(1.6\pm1.2) \times 10^{-2}$ erg~${\rm cm}^{-3}$, and the ratios of co-moving electron to synchrotron radiation energy density from $\sim1.89$ to $\sim59$ with an average value of $\sim10.4\pm4.3$. The number of electrons ranges from $\sim1\times10^{50}$ to $\sim1.9\times10^{51}$ with an average value of $\sim(9.3\pm1.6)\times10^{50}$, the fractional variability parameter $F_{\rm var}$ from 0.11 to 0.45 with an average value of $0.24\pm$0.02, the average flux of each sub-flare from 7.5 mJy to 13.7 mJy with an average value across all sub-flares of 10.34$\pm$0.5 mJy, and the average spectral index of each sub-flare from 1.24 to 1.54 with an average value across all sub-flares of 1.33$\pm$0.02. There are no significant correlations between $k$ (or $F_{\rm var}$) and one of the parameters ($R$, $B'$, $\gamma_{\rm e}$, $u_{\rm syn}'$, $u_{\rm e}'$, $u_{\rm e}'/u_{\rm syn}'$ and $N_{\rm e}$).

(iv) The flux variability in the long-term trend could be tracked by the variations of Doppler factor. The sub-flares superimposed in the long-term trend may be related with shock/turbulence in jet. For same emission zone (blob), the zone propagates along a bent trajectory or helical path in the jet, and produces the flares when the blob crosses the shock front/turbulence shell; for different zones, the flares are produced by a moving shock propagating along a bent trajectory and/or the shock encountering a region of turbulence. Such a scenario is likely to explain the flux variability in the short-term and long-term timescales.

\begin{acknowledgements}
We sincerely thank the referee for valuable comments and suggestions. We acknowledge the support of the staff from the Lijiang 2.4m (K. X. Lu, C. J. Wang, B. L. Lun,  Y. X. Xin, L. Xu, X. G. Yu, Y. F. Fan, K. Ye, S. S. He, X. Ding and D. Q. Wang) and BOOTES-telescopes (A. J. Castro-Tirado and E. Fernandez-Garcia from IAA-CSIC). This work is financially supported by the National Nature Science Foundation of China (11703078, 11991051, U1738124, 11733001), the Chinese Western Young Scholars Program provided by CAS (Y7XB018001), and the Yunnan Province Foundation (2019FB004). This research has made use of the NASA/IPAC Extragalactic Database (NED), that is operated by Jet Propulsion Laboratory, California Institute of Technology, under contract with the National Aeronautics and Space Administration. Data from the Steward Observatory spectropolarimetric monitoring project were used. This program is supported by Fermi Guest Investigator grants NNX08AW56G, NNX09AU10G, NNX12AO93G, and NNX15AU81G.

Software: LMFIT (Newville et al. 2016), SciPy (Jones et al. 2001), NumPy (Walt et al. 2011), Matplotlib (Barrett et al. 2005), and Statsmodels (Seabold \& Perktold 2010).
\end{acknowledgements}

\startlongtable
\begin{deluxetable}{cccr}
\tablecaption{The observational log in the $r$ band\label{Table1}}
\tablewidth{\columnwidth}
\tablecolumns{4}
\tablehead{\colhead{Date} & \colhead{Number of data points} &
\colhead{Time spans(hours)} & \colhead{Time resolutions(mins)}}
\startdata
58067 & 5 & 1.04 & 6.13 \\
58069 & 4 & 0.41 & 6.13 \\
58071 & 5 & 1.14 & 6.13 \\
58074 & 23 & 3.18 & 6.13 \\
58075 & 26 & 8.60 & 6.13 \\
58076 & 35 & 6.29 & 6.13 \\
58077 & 32 & 4.25 & 3.13 \\
58078 & 77 & 10.74 & 3.13 \\
58079 & 45 & 5.10 & 3.13 \\
58080 & 30 & 3.15 & 3.13 \\
58081 & 28 & 2.29 & 3.13 \\
58082 & 42 & 3.93 & 3.13 \\
58083 & 3 & 0.10 & 3.13 \\
58084 & 12 & 1.01 & 3.13 \\
58085 & 8 & 0.42 & 3.13 \\
58087 & 47 & 9.03 & 3.13 \\
58088 & 54 & 5.47 & 3.13 \\
58089 & 49 & 8.07 & 3.13 \\
58090 & 87 & 9.02 & 1.80 \\
58091 & 168 & 9.54 & 1.80 \\
58092 & 139 & 6.46 & 1.80 \\
58093 & 174 & 9.38 & 1.80 \\
58094 & 62 & 4.97 & 1.80 \\
58097 & 176 & 9.62 & 1.80 \\
58098 & 121 & 5.19 & 1.80 \\
58100 & 107 & 6.86 & 3.13 \\
58101 & 28 & 1.92 & 3.13 \\
58102 & 116 & 9.98 & 3.13 \\
58103 & 100 & 7.69 & 3.13 \\
58104 & 72 & 4.96 & 3.13 \\
58105 & 13 & 0.68 & 3.13 \\
58107 & 59 & 5.41 & 3.13 \\
58108 & 43 & 3.60 & 3.13 \\
58110 & 86 & 5.94 & 3.13 \\
58111 & 38 & 2.84 & 3.13 \\
58112 & 56 & 4.70 & 3.13 \\
58114 & 84 & 7.54 & 3.13 \\
58115 & 19 & 0.79 & 1.63 \\
58116 & 34 & 1.76 & 1.63 \\
58117 & 35 & 1.36 & 1.63 \\
58118 & 204 & 9.80 & 1.63 \\
58119 & 7 & 0.82 & 1.63 \\
58122 & 107 & 9.06 & 1.63 \\
58124 & 180 & 7.51 & 1.63 \\
58125 & 124 & 6.31 & 1.63 \\
58127 & 245 & 8.50 & 1.63 \\
58128 & 71 & 5.38 & 1.63 \\
58132 & 41 & 2.71 & 1.63 \\
58133 & 24 & 1.84 & 1.63 \\
58134 & 59 & 3.26 & 1.63 \\
58135 & 141 & 4.33 & 1.63 \\
58136 & 80 & 2.78 & 1.63 \\
58137 & 89 & 4.22 & 1.55 \\
58138 & 241 & 9.37 & 1.55 \\
58139 & 149 & 5.17 & 1.55 \\
58144 & 46 & 1.51 & 1.55 \\
58145 & 124 & 8.61 & 1.55 \\
58146 & 176 & 5.99 & 1.55 \\
58147 & 128 & 8.29 & 1.55 \\
58148 & 53 & 4.90 & 1.55 \\
58149 & 8 & 0.31 & 1.55 \\
58150 & 37 & 2.55 & 1.55 \\
58155 & 36 & 2.95 & 1.55 \\
58156 & 9 & 1.60 & 1.55 \\
58157 & 6 & 0.57 & 3.10 \\
58158 & 42 & 2.92 & 1.55 \\
58159 & 25 & 2.01 & 1.55 \\
58160 & 22 & 0.70 & 1.55 \\
58161 & 10 & 0.60 & 1.55 \\
58162 & 22 & 7.71 & 1.55 \\
58164 & 71 & 4.46 & 1.55 \\
58165 & 46 & 2.27 & 1.55 \\
58167 & 29 & 8.82 & 1.55 \\
58168 & 23 & 2.22 & 1.55 \\
58169 & 50 & 7.70 & 1.55 \\
58170 & 117 & 3.94 & 1.55 \\
58171 & 104 & 5.37 & 1.55 \\
58172 & 2 & 0.03 & 1.57 \\
58173 & 66 & 4.40 & 1.55 \\
58174 & 39 & 3.21 & 1.55 \\
58176 & 80 & 4.69 & 1.55 \\
58177 & 120 & 5.40 & 1.55 \\
58178 & 114 & 5.10 & 1.55 \\
58179 & 18 & 8.38 & 1.55 \\
58181 & 16 & 2.53 & 3.30 \\
58182 & 46 & 3.26 & 3.30 \\
58184 & 115 & 10.04 & 3.30 \\
58185 & 84 & 4.86 & 3.30 \\
58186 & 49 & 8.55 & 3.30 \\
58192 & 43 & 5.48 & 3.30 \\
58194 & 40 & 1.25 & 1.63 \\
58195 & 28 & 4.65 & 1.47 \\
58196 & 101 & 9.50 & 1.47 \\
58197 & 41 & 1.10 & 1.47 \\
58198 & 86 & 4.07 & 1.47 \\
58199 & 88 & 3.24 & 1.47 \\
58200 & 41 & 2.07 & 1.47 \\
58201 & 19 & 1.05 & 1.47 \\
58203 & 79 & 2.36 & 1.47 \\
58205 & 29 & 0.74 & 1.47 \\
58206 & 44 & 1.58 & 1.47 \\
58207 & 57 & 2.14 & 1.47 \\
58208 & 60 & 3.53 & 1.47 \\
58209 & 48 & 5.01 & 3.13 \\
58210 & 2 & 0.05 & 3.13 \\
58211 & 26 & 1.99 & 3.13 \\
58212 & 14 & 1.36 & 3.13 \\
58213 & 2 & 0.05 & 3.13 \\
58216 & 4 & 0.16 & 3.13 \\
58217 & 24 & 1.73 & 3.13 \\
58220 & 18 & 1.26 & 3.13 \\
58221 & 24 & 2.40 & 3.13 \\
58223 & 2 & 0.05 & 3.13 \\
58224 & 25 & 1.83 & 3.13 \\
58225 & 25 & 1.93 & 3.13 \\
58228 & 8 & 0.63 & 3.13 \\
58230 & 22 & 2.20 & 3.13 \\
58231 & 33 & 2.30 & 3.13 \\
58232 & 25 & 2.51 & 3.13 \\
58233 & 27 & 2.35 & 3.13 \\
58238 & 2 & 0.04 & 2.58 \\
58242 & 2 & 0.04 & 2.17 \\
58249 & 2 & 0.04 & 2.20 \\
58252 & 2 & 0.04 & 2.20 \\
58253 & 2 & 0.04 & 2.20 \\
58433 & 43 & 1.49 & 1.80 \\
58434 & 20 & 1.92 & 1.80 \\
58435 & 42 & 7.30 & 1.80 \\
58437 & 51 & 2.01 & 1.80 \\
58438 & 11 & 0.43 & 1.80 \\
58439 & 54 & 7.92 & 1.13 \\
58461 & 14 & 0.40 & 1.20 \\
58462 & 34 & 2.01 & 3.17 \\
58463 & 27 & 2.12 & 3.17 \\
58464 & 62 & 4.03 & 3.17 \\
58465 & 31 & 2.46 & 3.17 \\
58474 & 20 & 1.22 & 3.17 \\
58476 & 45 & 2.80 & 3.17 \\
58477 & 86 & 6.61 & 3.17 \\
58478 & 57 & 4.12 & 3.17 \\
58479 & 79 & 6.30 & 3.17 \\
58480 & 16 & 1.48 & 3.17 \\
58482 & 31 & 2.65 & 3.17 \\
58483 & 52 & 3.44 & 3.17 \\
58484 & 38 & 3.02 & 3.17 \\
58486 & 83 & 6.17 & 3.15 \\
58487 & 42 & 3.01 & 3.15 \\
58488 & 5 & 0.26 & 3.15 \\
58489 & 32 & 2.16 & 3.15 \\
58490 & 26 & 3.58 & 3.15 \\
58493 & 4 & 0.51 & 3.17 \\
58495 & 10 & 0.77 & 3.15 \\
58501 & 5 & 0.26 & 3.83 \\
58502 & 7 & 0.32 & 3.15 \\
58503 & 16 & 0.90 & 3.15 \\
58506 & 37 & 4.75 & 3.78 \\
58507 & 50 & 8.25 & 3.82 \\
58508 & 52 & 8.26 & 3.82 \\
58509 & 74 & 8.45 & 3.78 \\
58510 & 30 & 2.35 & 3.80 \\
58528 & 36 & 6.81 & 3.98 \\
58529 & 30 & 6.72 & 3.98 \\
58530 & 5 & 0.34 & 4.02 \\
58531 & 15 & 6.75 & 4.97 \\
58535 & 12 & 4.23 & 3.93 \\
58553 & 32 & 2.80 & 4.20 \\
58555 & 29 & 2.84 & 4.18 \\
58557 & 19 & 1.55 & 4.13 \\
58559 & 30 & 2.03 & 4.12 \\
58560 & 3 & 1.24 & 4.72 \\
58561 & 9 & 0.85 & 4.15 \\
58562 & 16 & 1.35 & 4.13 \\
58564 & 10 & 0.71 & 4.23 \\
58566 & 11 & 2.34 & 4.25 \\
58569 & 11 & 0.72 & 4.18 \\
58570 & 43 & 3.00 & 2.63 \\
58575 & 4 & 0.21 & 4.18 \\
58584 & 7 & 0.43 & 4.25 \\
58586 & 20 & 1.75 & 4.12 \\
58587 & 25 & 1.84 & 4.17 \\
58593 & 2 & 0.07 & 4.32 \\
58606 & 4 & 0.21 & 4.22 \\
58607 & 1 & --- & --- \\
58610 & 26 & 4.16 & 4.18 \\
58613 & 18 & 1.61 & 4.17 \\
58614 & 8 & 0.50 & 4.22 \\
58615 & 10 & 0.84 & 4.13 \\
58616 & 14 & 1.06 & 4.18 \\
58617 & 11 & 1.00 & 4.15 \\
58618 & 6 & 0.44 & 4.27 \\
58619 & 3 & 0.14 & 4.30 \\
58621 & 11 & 0.78 & 4.20 \\
58623 & 9 & 0.59 & 4.25 \\
58624 & 3 & 0.29 & 4.30 \\
58625 & 10 & 0.63 & 4.15 \\
58626 & 7 & 2.10 & 4.22 \\
58632 & 19 & 1.56 & 4.17 \\
58633 & 8 & 0.50 & 4.18 \\
58634 & 5 & 0.35 & 4.17 \\
58639 & 28 & 2.73 & 4.07 \\
58640 & 3 & 0.14 & 4.17 \\
\enddata
\tablecomments{The columns stand for observational date in the unit of MJD, number of data points, time spans per night in the unit of hours, time resolutions in the unit of minutes, respectively.}
\end{deluxetable}

\startlongtable
\begin{deluxetable}{cccccccccccr}
\tablecaption{The observational data\label{Table2}}
\tablecolumns{12}
\tablewidth{0pt}
\centering
\tablehead{\colhead{$g$-date} & \colhead{$g$-mag} & \colhead{$g$-err1} & \colhead{$g$-err2} & \colhead{$r$-date} & \colhead{$r$-mag} & \colhead{$r$-err1} & \colhead{$r$-err2} & \colhead{$i$-date} & \colhead{$i$-mag} & \colhead{$i$-err1} & \colhead{$i$-err2}
}
\startdata
58067.8947	&	14.0342	&	0.007	&	0.0104	&	58067.8961	&	13.5978	&	0.0082	&	0.0094	&	58067.8975	&	13.4146	&	0.0096	&	0.0174	\\
58067.9251	&	14.0502	&	0.007	&	0.0104	&	58067.9265	&	13.5943	&	0.0082	&	0.0094	&	58067.9279	&	13.3926	&	0.0086	&	0.0174	\\
58067.9294	&	14.0406	&	0.007	&	0.0104	&	58067.9308	&	13.5823	&	0.0082	&	0.0094	&	58067.9322	&	13.3986	&	0.0096	&	0.0174	\\
58067.9336	&	14.0346	&	0.0072	&	0.0104	&	58067.935	&	13.6088	&	0.0082	&	0.0094	&	58067.9364	&	13.3931	&	0.0096	&	0.0174	\\
58067.9379	&	14.0472	&	0.0082	&	0.0104	&	58067.9393	&	13.5768	&	0.0094	&	0.0094	&	58067.9407	&	13.3886	&	0.0108	&	0.0174	\\
58069.9099	&	14.1806	&	0.0054	&	0.0067	&	58069.9113	&	13.6628	&	0.007	&	0.0082	&	58069.9128	&	13.4751	&	0.0082	&	0.0147	\\
58069.9185	&	14.1422	&	0.0045	&	0.0067	&	58069.9199	&	13.6693	&	0.007	&	0.0082	&	58069.9213	&	13.4741	&	0.0082	&	0.0147	\\
58069.9227	&	14.1466	&	0.0056	&	0.0067	&	58069.9241	&	13.6653	&	0.0078	&	0.0082	&	58069.9255	&	13.4686	&	0.0094	&	0.0147	\\
58069.927	&	14.1166	&	0.0047	&	0.0067	&	58069.9284	&	13.6503	&	0.0068	&	0.0082	&	58069.9298	&	13.4631	&	0.0072	&	0.0147	\\
58071.8953	&	14.2196	&	0.0371	&	0.043	&	58071.8967	&	13.6648	&	0.0113	&	0.0359	&	58071.8981	&	13.3726	&	0.0194	&	0.0419	\\
\enddata
\tablecomments{Column (1) is the modified Julian day (MJD) of the observation in the $g$ band, column (2) the magnitude in the $g$ band, column (3) the Poisson errors from the target source and the comparison stars measured by IRAF, column (4) the rms errors. The meanings from column (5) to column (12) are same as that from column (1) to column (4) except for the different bands.
This table is available in its entirety in machine-readable form.}
\end{deluxetable}

\begin{longrotatetable}
\begin{deluxetable}{ccccccccccccccccr}
\tablecaption{The fitting results\label{Table3}}
\tablecolumns{17}
\tablewidth{0pt}
\centering
\tabletypesize{\tiny}
\tablehead{\colhead{MJD} & \colhead{$r$} & \colhead{$k$} & \colhead{$F_{\rm var}$} & \colhead{$<F>$} & \colhead{$<\alpha>$} & \colhead{$t_{\rm 0}$} & \colhead{$t_{\rm r}$} & \colhead{$t_{\rm d}$} & \colhead{$F_{\rm peak}$} & \colhead{$R$} & \colhead{$B'$} & \colhead{$\gamma_{\rm e}$} & \colhead{$u'_{\rm syn}$} & \colhead{$N_{\rm e}$} & \colhead{$u'_{\rm e}$} & \colhead{$u'_{\rm e}/u'_{\rm syn}$}
}
\startdata
58079-58100$^{(1)}$	&	0.83	&	0.067$\pm$0.001	&	0.218$\pm$0.001	&	11.15$\pm$0.07	&	1.507$\pm$0.006	&	58083.75$\pm$0.39	&	3.3$\pm$0.31	&	7.28$\pm$0.29	&	11.5	&	0.023	&	0.070	&	14855	&	0.68	&	1.14	&	9.16	&	13.47	\\
58100-58119$^{(2)}$	&	0.58	&	0.019$\pm$0.001	&	0.257$\pm$0.001	&	14.53$\pm$0.11	&	1.241$\pm$0.003	&	58116.14$\pm$0.21	&	4.22$\pm$0.15	&	2.83$\pm$0.21	&	18	&	0.029	&	0.131	&	10842	&	0.65	&	0.95	&	2.67	&	4.10	\\
58119-58150$^{(3)}$	&	0.41	&	0.0055$\pm$0.0001	&	0.209$\pm$0.001	&	18.54$\pm$0.09	&	1.219$\pm$0.001	&	58130.08$\pm$0.19	&	3.92$\pm$0.18	&	6.17$\pm$0.18	&	23	&	0.027	&	0.078	&	14058	&	0.96	&	2.04	&	9.27	&	9.61	\\
58150-58180$^{(4)}$	&	0.72	&	0.012$\pm$0.0001	&	0.197$\pm$0.001	&	23.92$\pm$0.14	&	1.172$\pm$0.002	&	58165.15$\pm$0.4	&	7.47$\pm$0.28	&	9.98$\pm$0.4	&	26	&	0.052	&	0.056	&	16502	&	0.30	&	3.17	&	2.45	&	8.16	\\
58180-58235$^{(5)}$	&	0.36	&	0.0061$\pm$0.0001	&	0.192$\pm$0.001	&	26.37$\pm$0.14	&	1.201$\pm$0.002	&	58215.92$\pm$0.5	&	11.97$\pm$0.42	&	10.83$\pm$0.38	&	38	&	0.084	&	0.053	&	16958	&	0.17	&	4.90	&	0.94	&	5.53	\\
58430-58496$^{(6)}$	&	0.71	&	0.017$\pm$0.001	&	0.304$\pm$0.001	&	15.04$\pm$0.14	&	1.355$\pm$0.004	&	58471.27$\pm$0.34	&	19.29$\pm$0.28	&	9.82$\pm$0.28	&	17.5	&	0.135	&	0.057	&	16414	&	0.03	&	2.11	&	0.09	&	3.11	\\
58496-58525$^{(7)}$	&	0.66	&	0.026$\pm$0.002	&	0.112$\pm$0.001	&	11.61$\pm$0.08	&	1.331$\pm$0.003	&	58508.51$\pm$0.62	&	3.26$\pm$0.47	&	2.53$\pm$0.65	&	7	&	0.023	&	0.141	&	10444	&	0.42	&	0.34	&	2.01	&	4.74	\\
58525-58538$^{(8)}$	&	0.72	&	0.046$\pm$0.005	&	0.184$\pm$0.001	&	8.84$\pm$0.17	&	1.313$\pm$0.011	&	58531.19$\pm$0.39	&	1.34$\pm$0.36	&	0.48$\pm$0.35	&	5.9$\pm$0.8	&	0.009	&	0.427	&	6001	&	2.12	&	0.10	&	4.63	&	2.19	\\
58538-58565$^{(9)}$	&	0.77	&	0.048$\pm$0.003	&	0.265$\pm$0.002	&	13.21$\pm$0.29	&	1.314$\pm$0.016	&	58557.57$\pm$0.25	&	9.01$\pm$0.73	&	2.92$\pm$0.25	&	9.5	&	0.063	&	0.128	&	10955	&	0.08	&	0.51	&	0.15	&	1.98	\\
58565-58617$^{(10)}$	&	0.52	&	0.011$\pm$0.001	&	0.446$\pm$0.002	&	13.00$\pm$0.41	&	1.325$\pm$0.009	&	58587.18$\pm$1.31	&	5.71$\pm$0.47	&	15.85$\pm$1.73	&	17$\pm$1.3	&	0.040	&	0.041	&	19254	&	0.34	&	2.82	&	5.69	&	16.95	\\
58617-58626$^{(11)}$	&	0.29	&	0.016$\pm$0.005	&	0.255$\pm$0.004	&	12.83$\pm$0.45	&	1.427$\pm$0.019	&	58621.91$\pm$0.27	&	1.80$\pm$0.3	&	0.76$\pm$0.15	&	8.5	&	0.013	&	0.314	&	6995	&	1.69	&	0.19	&	4.36	&	2.58	\\
58626-58645$^{(12)}$	&	0.58	&	0.014$\pm$0.002	&	0.182$\pm$0.002	&	19.67$\pm$0.43	&	1.209$\pm$0.01	&	58631.31$\pm$0.4	&	2.75$\pm$0.18	&	9.07$\pm$0.65	&	14.5	&	0.019	&	0.060	&	15985	&	1.23	&	1.66	&	24.88	&	20.14	\\
\enddata
\tablecomments{Column (1) is the time ranges of a single flare in MJD in which the numbers in the upper right corner represent the order numbers of sub-flares, column (2) the correlation coefficients of WLS fitting with all of the chance probability $P<10^{-4}$ except for MJD=58617-58626 having $P=0.02$, column (3) the slopes of WLS fitting, column (4) the fractional variability, column (5) the average flux in mJy, column (6) the average spectral index, column (7) the positions of peak in MJD, column (8) the rise time in days, column (9) the decay time in days, column (10) the amplitude of the flares in mJy, column (11) the upper limit of emitting region in pc, column (12) the lower limit of magnetic field strength in Gauss, column (13) the upper limit of random Lorentz factor, column (14) the lower limit of co-moving energy density of the synchrotron radiation in $10^{-3}$ erg~${\rm cm}^{-3}$, column (15) the number of electrons in $10^{51}$, column (16) the co-moving energy density of the electrons in $10^{-3}$ erg~${\rm cm}^{-3}$, column (17) the ratio of co-moving electron to synchrotron radiation energy density.}
\end{deluxetable}
\end{longrotatetable}

\begin{longrotatetable}
\begin{deluxetable}{ccccccccccccccccr}
\tablecaption{The fitting results of removing the long-term trend of 60-days binning data\label{Table3}}
\tablecolumns{17}
\tablewidth{0pt}
\centering
\tabletypesize{\tiny}
\tablehead{\colhead{MJD} & \colhead{$r$} & \colhead{$k$} & \colhead{$F_{\rm var}$} & \colhead{$<F>$} & \colhead{$<\alpha>$} & \colhead{$t_{\rm 0}$} & \colhead{$t_{\rm r}$} & \colhead{$t_{\rm d}$} & \colhead{$F_{\rm peak}$} & \colhead{$R$} & \colhead{$B'$} & \colhead{$\gamma_{\rm e}$} & \colhead{$u'_{\rm syn}$} & \colhead{$N_{\rm e}$} & \colhead{$u'_{\rm e}$} & \colhead{$u'_{\rm e}/u'_{\rm syn}$}
}
\startdata
58079-58100$^{(1)}$	&	0.83	&	0.095$\pm$0.002	&	0.235$\pm$0.001	&	7.53$\pm$0.05	&	1.542$\pm$0.006	&	58083.93$\pm$0.25	&	3.11$\pm$0.18	&	5.94$\pm$0.21	&	10.3	&	0.022	&	0.080	&	13881	&	0.69	&	0.89	&	8.00	&	11.66	\\
58100-58119$^{(2)}$	&	0.53	&	0.032$\pm$0.002	&	0.217$\pm$0.001	&	8.94$\pm$0.06	&	1.304$\pm$0.004	&	58116.25$\pm$0.17	&	3.89$\pm$0.13	&	2.42$\pm$0.15	&	12	&	0.027	&	0.145	&	10291	&	0.51	&	0.57	&	1.94	&	3.80	\\
58119-58150$^{(3)}$	&	0.47	&	0.011$\pm$0.0001	&	0.236$\pm$0.001	&	9.75$\pm$0.05	&	1.319$\pm$0.001	&	58130.13$\pm$0.18	&	3.83$\pm$0.15	&	5.44$\pm$0.17	&	14.3	&	0.027	&	0.085	&	13480	&	0.63	&	1.16	&	5.45	&	8.67	\\
58150-58180$^{(4)}$	&	0.76	&	0.029$\pm$0.001	&	0.202$\pm$0.001	&	10.53$\pm$0.07	&	1.293$\pm$0.002	&	58164.63$\pm$0.36	&	6.47$\pm$0.25	&	9.41$\pm$0.35	&	12	&	0.045	&	0.059	&	16182	&	0.18	&	1.41	&	1.64	&	8.88	\\
58180-58235$^{(5)}$	&	0.25	&	0.011$\pm$0.001	&	0.172$\pm$0.001	&	10.07$\pm$0.05	&	1.337$\pm$0.002	&	58215.72$\pm$0.48	&	10.4$\pm$0.36	&	7.94$\pm$0.36	&	15	&	0.073	&	0.066	&	15291	&	0.09	&	1.57	&	0.42	&	4.66	\\
58430-58496$^{(6)}$	&	0.61	&	0.016$\pm$0.001	&	0.32$\pm$0.001	&	13.65$\pm$0.14	&	1.328$\pm$0.004	&	58474.71$\pm$0.21	&	18.27$\pm$0.22	&	7.95$\pm$0.21	&	14.5	&	0.128	&	0.066	&	15298	&	0.03	&	1.52	&	0.07	&	2.66	\\
58496-58525$^{(7)}$	&	0.66	&	0.026$\pm$0.002	&	0.113$\pm$0.001	&	11.60$\pm$0.08	&	1.331$\pm$0.003	&	58508.43$\pm$0.58	&	4.37$\pm$0.47	&	3.31$\pm$0.68	&	7	&	0.031	&	0.118	&	11423	&	0.24	&	0.41	&	1.09	&	4.63	\\
58525-58538$^{(8)}$	&	0.73	&	0.047$\pm$0.005	&	0.185$\pm$0.001	&	8.67$\pm$0.16	&	1.306$\pm$0.01	&	58531.29$\pm$0.27	&	1.58$\pm$0.32	&	0.49$\pm$0.27	&	6.1$\pm$0.7	&	0.011	&	0.421	&	6043	&	1.57	&	0.10	&	2.98	&	1.89	\\
58538-58565$^{(9)}$	&	0.76	&	0.051$\pm$0.004	&	0.274$\pm$0.002	&	11.96$\pm$0.27	&	1.292$\pm$0.016	&	58556.60$\pm$0.29	&	6$\pm$0.2	&	3$\pm$0.25	&	9	&	0.042	&	0.126	&	11055	&	0.16	&	0.49	&	0.49	&	3.05	\\
58565-58617$^{(10)}$	&	0.58	&	0.017$\pm$0.002	&	0.453$\pm$0.002	&	10.06$\pm$0.32	&	1.307$\pm$0.009	&	58588.71$\pm$1.08	&	5.48$\pm$0.46	&	11.42$\pm$1.02	&	14$\pm$0.96	&	0.038	&	0.052	&	17261	&	0.30	&	1.87	&	3.82	&	12.73	\\
58617-58626$^{(11)}$	&	0.32	&	0.025$\pm$0.011	&	0.255$\pm$0.004	&	8.72$\pm$0.31	&	1.437$\pm$0.019	&	58621.78$\pm$0.28	&	1.55$\pm$0.28	&	0.8$\pm$0.15	&	6.1	&	0.011	&	0.304	&	7115	&	1.64	&	0.14	&	5.15	&	3.15	\\
58626-58645$^{(12)}$	&	0.59	&	0.021$\pm$0.003	&	0.187$\pm$0.002	&	12.57$\pm$0.28	&	1.247$\pm$0.01	&	58628.71$\pm$1.3	&	1.3$\pm$0.64	&	12.55$\pm$2.2	&	6.9	&	0.009	&	0.048	&	17812	&	2.63	&	0.98	&	155.05	&	58.96	\\
\enddata
\tablecomments{The meanings of the columns are same as that of Table 3.}
\end{deluxetable}
\end{longrotatetable}

\end{document}